\documentclass[showpacs,aps,prl,reprint,superscriptaddress]{revtex4-1}
\usepackage{amsmath}
\usepackage{amsfonts}
\usepackage{amssymb}
\usepackage{amsthm}
\usepackage{graphicx}
\usepackage{CJK}
\usepackage{color}
\usepackage{array}
\usepackage{times}
\usepackage{mathptmx}
\usepackage[colorlinks, linkcolor=blue, urlcolor=blue, anchorcolor=blue, citecolor=blue]{hyperref}

\begin{document}
\title{Spin-mixing-tunneling network model for Anderson transitions in two-dimensional disordered spinful electrons}
\author{Jie Lu}
\email{jlu@hebtu.edu.cn}
\affiliation{College of Physics, Hebei Advanced Thin Films Laboratory, Hebei Normal University, Shijiazhuang 050024, China}
\author{Mei Li}
\affiliation{School of Physics and Technology, Center for Electron Microscopy and MOE Key Laboratory of Artificial Micro- and Nano-structures, Wuhan University, Wuhan 430072, China}
\author{Bin Xi}
\affiliation{College of Physics Science and Technology, Yangzhou University, Yangzhou 225002, People’s Republic of China}
\date{\today}

\begin{abstract}
We consider Anderson transitions in two-dimensional spinful electron gases
subject to random scalar potentials with time-reversal-symmetric 
spin-mixing tunneling (SMT) and spin-preserving tunneling (SPT)
at potential saddle points (PSPs).
A symplectic quantum network model, named as SMT-QNM, is constructed in which 
SMT and SPT have the same status and contribute independent tunneling channels 
rather than sharing a total-probability-fixed one. 
Two-dimensional continuous Dirac Hamiltonian is then extracted out from
this discrete network model as the generator of certain unitary transformation.
With the help of high-accuracy numerics based on transfer matrix technique,
finite-size analysis on two-terminal conductance and normalized localization length
provides a phase diagram drawn in the SMT-SPT plane.
As a manifestation of symplectic ensembles, a normal-metal (NM) phase emerges between
the quantum spin Hall (QSH) and normal-insulator (NI) phases when SMT appears.
We systematically analyze the quantum phases on the boundary and in the interior
of the phase space.
Particularly, the phase diagram is closely related to that of
disordered three-dimensional weak topological insulators by appropriate
parameter mapping.
At last, if time-reversal symmetry in electron trajectories between PSPs
is destroyed, the system falls into unitary class with no more NM phase.
A direct SMT-driven transition from QSH to NI phases exists and can be explained
by spin-flip backscattering between the degenerate doublets at the same sample edge.
\end{abstract}


\maketitle


\section{I. Introduction}
Anderson transitions (ATs), i.e., transitions between localized and delocalized 
quantum phases in disordered electronic systems, have attracted intense and 
continuous attention since its proposal\cite{Anderson_1958} due to its fundamental 
significance in condensed matter
physics\cite{Lee_1985,Kramer_1993,Huckestein_1995,Evers_Mirlin_2008}.
In 1970s and 1980s, scaling-theory and field-theory approaches revealed the
connections between Anderson transition and conventional second-order
phase transitions\cite{Lee_1985,Kramer_1993,Huckestein_1995}.
In 1990s, the symmetry classification of disordered systems was achieved based on 
its relation to the classical symmetric spaces\cite{Zirnbauer_1996,Zirnbauer_1997_prb,Caselle_2004}.
Later, the completeness of this classification is proved in 2005\cite{Heinzner_2005}.
Now we know there are totally ten symmetry classes according to how many discrete 
symmetries are obeyed by the underlying physical system.
When a system only has symmetries translationally invariant in
energy, such as the time-reversal symmetry (TRS) and spin-rotation symmetry (SRS),
it falls into one of the three traditional Wigner-Dyson
classes (unitary, orthogonal and symplectic)\cite{Wigner_1951,Dyson_1962}.
However, if we focus on some particular value of energy, extra discrete symmetries 
could arise and lead to novel symmetry classes.
In condensed matter systems described by tight-binding models on a bipartite lattice 
with randomness only residing in hopping terms, three chiral classes are
identified\cite{Zirnbauer_1996}.
The remaining four were discovered in superconducting systems and 
known as the Bogoliubov-de Gennes classes\cite{Zirnbauer_1997_prb}.
In the past decades, ATs in these ten classes have been investigated intensively 
and considerable progress has been made in various directions, such as their
scaling-theory and field-theory descriptions\cite{Lee_1985,Kramer_1993,Huckestein_1995},
multifractality in critical wave functions\cite{Janssen_1994,Mudry_1996,Evers_Mirlin_2000,
	Evers_2001,Obuse_2004_prb,Obuse_2007_prl,Obuse_2010_prb_1}
and level statistics at criticality\cite{Mirlin_2000,XiongGang_2001_PRL,XiongGang_2006_JPCM,Obuse_2005_prb,GarciaGarcia_2007},
etc.

Recently, the spin-orbit-induced topological materials, named as topological 
insulators (TIs), have received intensive attention\cite{RMP_2010_Hasan_Kane,RMP_2011_XLQi_SCZhang,RMP_2015_Beenakker,RPP_2016_ZHQiao_QNiu,
RMP_2016_Bansil,RMP_2016_Witten,RMP_2016_Ryu}.
In TIs, the interplay between topology and symmetry greatly enriches our knowledge 
of quantum states\cite{RMP_2016_Ryu,Fu_2012_prl,Fulga_2012_prb,Ryu_2007_prl,Ryu_2010_njp}.
First, the TRS is crucial for their realization and stabilization.
Second, the spin-orbit coupling (SOC) destroys the SRS, thus makes TIs belong to 
the Wigner-Dyson symplectic class.
In two dimensions (2D), they are the well-known quantum spin Hall (QSH) ensembles.
In disordered QSH systems, ATs can be extended from traditional metal-insulator 
transitions to a broader sense which includes transition between
topologically trivial and nontrivial phases\cite{Evers_Mirlin_2008}.
In the past decade, great efforts have been devoted into this issue\cite{Ryu_2010_njp,
Obuse_2007_prb,Obuse_2008_prb,Obuse_2010_prb_2,Obuse_2014_prb,Slevin_2012,Ryu_2012_prb,
Slevin_2014_njp,XRWang_2015_PRL,CWang_2017_PRB}.
The widely-used framework is to construct a quantum network model which 
consists of two copies of Chalker-Coddington random network model (CC-RNM) 
describing up and down spins, as well as certain coupling describing spin-flip process.
If spin flip occurs in electron trajectories between potential saddle points (PSPs), 
it is the well-known spin-orbit coupling (SOC).
While if it takes place at the PSPs, it is the spin-mixing tunneling (SMT) 
which is the main focus in this work.
Recently, a $Z_2$ quantum network model ($Z_2$-QNM) is
proposed\cite{Ryu_2010_njp,Obuse_2007_prb,Obuse_2008_prb,Obuse_2010_prb_2,Obuse_2014_prb,Slevin_2012}
in which SMT at PSPs are considered.
It belongs to the Wigner-Dyson symplectic class and a series of work declare that it
provides a good description of ATs in 2D disordered spinful electron gases(2D-DSEGs).
In $Z_2$-QNM, at PSPs the total tunneling probability are fixed, which means
SMT takes away part of the probability from the spin-preserving tunneling (SPT) process.
However, from the basic principles of quantum tunneling SMT
provides an additional channel and should not affect the existing SPT.
In this work, we treat the SMT as an independent quantum tunneling channel and build a new
network model, namely the ``SMT-QNM", to provide an alternative perspective to understand
ATs in 2D-DSEGs.

This paper is organized as follows.
In Sec. II the SMT-QNM is systematically built up based on probability 
conservation and TRS at PSPs.
Then the 2D continuous Dirac Hamiltonian with ``valley" degree of freedom
is extracted out.
In Sec. III numerical algorithms using transfer matrix technique
for finite-size analysis on two-terminal conductance and normalized 
localization length are reviewed.
Based on them, in Sec. IV the quantum phases of SMT-QNM are investigated
and a phase diagram is then obtained.
We discuss its close connection with that of the disordered 3D weak TIs.
In Sec. V, we consider the case when TRS in electron trajectories between PSPs
is destroyed. The system then falls into unitary class.
We briefly summarize the quantum phases and phase transitions therein.
Finally, the concluding remarks are provided in the last section.

\section{II. The SMT-QNM}
\subsection{II.A Brief review of CC-RNM}
Under a strong magnetic field $\vec{B}=B\hat{z}$, the motion of an electron in a 
smooth enough 2D random scalar potential $V(\vec{r})$ can be decomposed into a 
rapid cyclotron gyration and a slow drift of the guiding center along an 
equipotential contour which is generally composed of numerous loops around
potential valleys or peaks\cite{Chalker_Coddington_1988,Kramer_2005}.
The drifting direction of electrons in each loop is uni-directional (chiral):
$\vec{v}(\vec{r})=\nabla V(\vec{r})\times\vec{B}/(eB^2)$.
At PSPs, electrons' reflecting along equipotential lines and their mutual tunneling
are the essential physical ingredients for constructing a network model 
describing quantum criticality in disordered 2D systems.
For modelization, the PSPs are arranged to form a 2D square lattice with
the interconnected links representing electron flows along equipotential lines.
The potential peaks and valleys distribute alternatively in the square plaquettes 
enclosed by the links. This endues definite propagating direction of electron 
flows on the links and then divides the PSPs into two subgroups: 
the S- and S'-types (see Fig. 1a and 1b).
At each PSP, two incoming and two outgoing electron flows intersect hence 
lead to a $2\times2$ scattering matrix.
Quantum tunneling only occurs at PSPs and in the simplest case can be assumed identical.
At last, disorder is introduced by random phases along links.
This is the basic framework of CC-RNM.
In all illustration figures in this paper, we adopt the following sketch rules: 
if $r>t$, the reflecting (tunneling) routes are depicted by solid (dash) curves
and vice versa.

For a S-type PSP at $\mathbf{R}$, its scattering matrix is,
\begin{equation}\label{Scattering_matrix_S_0_CCRNM}
\left(
\begin{matrix}
Z_{2}^{\mathrm{o}} \\
Z_{4}^{\mathrm{o}} \\
\end{matrix}
\right)
=s_{\mathbf{R}}^{\mathrm{CC}}
\left(
\begin{matrix}
Z_{1}^{\mathrm{i}} \\
Z_{3}^{\mathrm{i}} \\
\end{matrix}
\right),\quad s_{\mathbf{R}}^{\mathrm{CC}}=\Psi_{\mathbf{R}}^{24}S_{\mathrm{CC}}\Psi_{\mathbf{R}}^{13}.
\end{equation}
where $Z_{j}^{\mathrm{o(i)}}$ is the outgoing (incoming) electron flow amplitude at link $j$,
$\Psi_{\mathbf{R}}^{jk}\equiv \mathrm{diag}(e^{\mathrm{i}\psi_{\mathbf{R}}^{j}},
e^{\mathrm{i}\psi_{\mathbf{R}}^{k}})$ is a diagonal matrix,
with $\psi_{\mathbf{R}}^{j}$ being the dynamical phase an electron acquires 
when propagating on link $j$ between the observation point and the PSP at $\mathbf{R}$.
The kernel matrix $S_{\mathrm{CC}}$ has the general form,
\begin{equation}\label{Scattering_matrix_S_1_CCRNM}
S_{\mathrm{CC}}=\left(
\begin{matrix}
r & t \\
\eta_t t & \eta_r r \\
\end{matrix}
\right),
\end{equation}
where $r=\sqrt{p}$ ($t=\sqrt{1-p}$) measuring the reflecting (tunneling) 
amplitude at a PSP,
and $p$ is related to the Fermi level of the system\cite{Kramer_2005}. 
$\eta_{t(r)}$ are undetermined coefficients.
In steady states, probability conservation at any PSP requires $\eta_{t(r)}=e^{\mathrm{i}\phi_{t(r)}}$
and $|\phi_t-\phi_r|=(2n+1)\pi$. Clearly,
\begin{equation}\label{TRS_broken_CCRNM}
\left(\mathrm{i}\sigma_{y}\right)
S_{\mathrm{CC}}^{*}
\left(-\mathrm{i}\sigma_{y}\right)
=\eta_r^{-1}S_{\mathrm{CC}} \neq S_{\mathrm{CC}}^{\dagger},
\end{equation}
which means TRS is broken thus the CC-RNM belongs to the unitary class.
Throughout this work, $\eta_r=-\eta_t=-1$ which is also the choice in most literatures.

\subsection{II.B Scatter matrices of SMT-QNM}
To describe ATs in 2D-DSEGs, the CC-RNM should be generalized to include spins,
providing the following hypotheses.
First, the potential profile is identical for any spin orientation.
Second, the absence of external magnetic fields makes TRS possible
which turns the original uni-directed electron flow on each link to a Kramers doublet.
Opposite spin components then ``feel" opposite effective magnetic fields,
forming two copies of CC-RNM with opposite chirality.
Third, appropriate coupling should be introduced between the two copies of
CC-RNM to describe spin-flip process.
Generally, spin flip can occur anywhere.
In real modelization, two strategies are most common: 
(a) it only occurs on the links between PSPs; (b) it only occurs at the PSPs.
The first strategy reflects the SOC while the second one is the SMT.

\begin{figure}[htbp]
	\centering
	\scalebox{0.5}[0.5]{\includegraphics[width=0.92\textwidth]{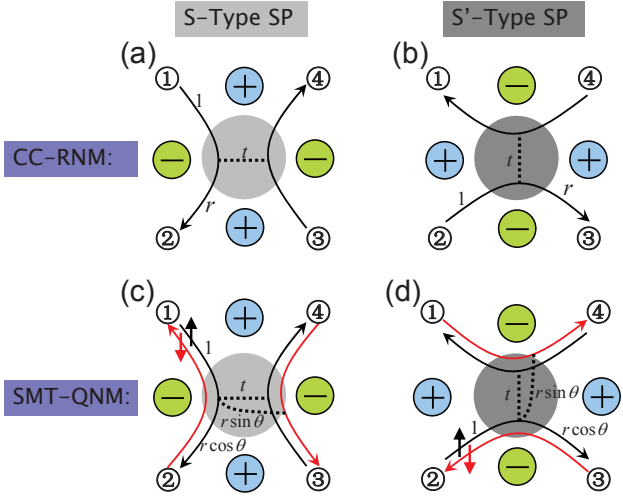}}
	\caption{(Color online) Schematics of CC-RNM and SMT-QNM. (a) and (b) show
		the S- and S'-type PSPs in CC-RNM. At each PSP, two incoming and two
		outgoing electron flows intersect with tunneling amplitude $\sqrt{1-p}$.
		Blue (green) circles with ``$+(-)$" inside denote the potential peaks (valleys).
		(c) and (d) show the counterparts of (a) and (b) in SMT-QNM where the spin degree
		of freedom is included. The original chiral electron flow on each link
		is generalized to a Kramers doublet. Throughout this paper, black (red)
		means spin-up (-down). In addition, at each PSP a SMT with amplitude 
		``$\sin\theta$" is introduced.}
	\label{fig01}
\end{figure}

The $Z_2$-QNM proposed
in Refs.\cite{Ryu_2010_njp,Obuse_2007_prb,Obuse_2008_prb,Obuse_2010_prb_2,Obuse_2014_prb}
follows the second strategy, however views SMT and SPT as two competing processes
sharing a fixed probability ``$t^2$".
In this work, the SPT channel remains unperturbed. Meantime we treat SMT as an independent 
quantum tunneling channel and construct the SMT-QNM to understand ATs in 2D-DSEGs.
For S-type PSPs (See Fig. 1c), the scattering matrix at position $\mathbf{R}$ reads,
\begin{equation}\label{Scattering_matrix_S_0_SMTQNM}
\left(
  \begin{matrix}
    Z_{2\uparrow}^{\mathrm{o}} \\
    Z_{1\downarrow}^{\mathrm{o}} \\
    Z_{4\uparrow}^{\mathrm{o}} \\
    Z_{3\downarrow}^{\mathrm{o}} \\
  \end{matrix}
\right)
=s_{\mathbf{R}}^{\mathrm{SMT}}
\left(
  \begin{matrix}
    Z_{1\uparrow}^{\mathrm{i}} \\
    Z_{2\downarrow}^{\mathrm{i}} \\
    Z_{3\uparrow}^{\mathrm{i}} \\
    Z_{4\downarrow}^{\mathrm{i}} \\
  \end{matrix}
\right),\quad s_{\mathbf{R}}^{\mathrm{SMT}}=\Psi_{\mathbf{R}}^{2143}S_{\mathrm{SMT}}\Psi_{\mathbf{R}}^{1234}.
\end{equation}
where $Z_{j\sigma}^{\mathrm{o(i)}}$ is the outgoing (incoming) electron flow amplitude at
link $j$ with spin $\sigma (\uparrow \mathrm{or} \downarrow)$,
$\Psi_{\mathbf{R}}^{jklm}\equiv \mathrm{diag}(e^{\mathrm{i}\psi_{\mathbf{R}}^{j}},
e^{\mathrm{i}\psi_{\mathbf{R}}^{k}},
e^{\mathrm{i}\psi_{\mathbf{R}}^{l}},
e^{\mathrm{i}\psi_{\mathbf{R}}^{m}})$ with
$\psi_{\mathbf{R}}^{j}$ representing the phase an electron acquires when
propagating on link $j$ between the observation point and the PSP at $\mathbf{R}$.
We have neglected the spin index since the Kramers pair of electron flows
have the same accumulated phase on the same link.
To mimic the randomness in PSP distribution, these phases are distributed 
uniformly and independently in the region $[0,2\pi)$.
If we focus on the very point where a PSP locates, $\Psi_{\mathbf{R}}^{jklm}$
then becomes unity.
The kernel matrix $S_{\mathrm{SMT}}$ describes the reflecting and tunneling 
at a general S-type PSP and has the following structure,
\begin{equation}\label{Scattering_matrix_S_1_SMTQNM}
  S_{\mathrm{SMT}}=\left(
               \begin{matrix}
                 \left(
                 \begin{matrix}
                 r_1 & 0 \\
                 0 & r_1^{*} \\
                 \end{matrix}
                 \right) & Q \\
                 Q^{\dagger} &
                 \left(
                 \begin{matrix}
                 -r_1 & 0 \\
                 0 & -r_1^{*} \\
                 \end{matrix}
                 \right)
                 \\
               \end{matrix}
             \right),
\end{equation}
where ``$\dagger$" means matrix complex conjugate.
For this scattering matrix, several points need to be clarified.
First, it is hermitian due to TRS. Second, $|r_1|\le r$ since
\emph{SMT is an additional tunneling channel hence takes probability away from 
	reflecting rather than SPT process}.
For simplicity, $r_1$ can be defined as $r_1=r\cos\theta$ (thus is real),
with $\theta\in \left[0,\pi/2\right]$ describing the strength of SMT.
Third, probability conservation in steady states at any PSP requires
the scattering matrix to be unitary,
\begin{eqnarray}\label{S_ProbabilityConservation_SMTQNM}
S_{\mathrm{SMT}}^{\dagger}S_{\mathrm{SMT}}=S_{\mathrm{SMT}}S_{\mathrm{SMT}}^{\dagger}=\sigma_0\otimes\sigma_0,
\end{eqnarray}
which gives
\begin{equation}\label{Q_definition_0_SMTQNM}
QQ^{\dagger}=Q^{\dagger}Q=\left(t^2+r^2\sin^2\theta\right)\sigma_0,
\end{equation}
where $\sigma_0$ is the $2\times2$ unit matrix.
Fourth, TRS requires
\begin{equation}\label{S_TRS_SMTQNM}
\left(
\begin{matrix}
\mathrm{i}\sigma_y & 0 \\
0 & \mathrm{i}\sigma_y \\
\end{matrix}
\right) S_{\mathrm{SMT}}^{*}
\left(
\begin{matrix}
-\mathrm{i}\sigma_y & 0 \\
0 & -\mathrm{i}\sigma_y \\
\end{matrix}
\right)
=S_{\mathrm{SMT}}^{\dagger},
\end{equation}
where $\sigma_{x,y,z}$ are the Pauli matrices.
This gives,
\begin{equation}\label{Q_definition_1_SMTQNM}
Q=\sigma_y Q^{*} \sigma_y.
\end{equation}

By writing $Q$ as
\begin{equation}\label{Q_definition_2_SMTQNM}
\begin{array}{c}
Q=a_0\sigma_0+ \sum_{k} a_k\sigma_k,\quad a_0,a_{k=x,y,z}\in \mathbb{C},
\end{array}
\end{equation}
Eq. (\ref{Q_definition_0_SMTQNM}) turns to
\begin{equation}\label{Q_definition_3_SMTQNM}
\begin{array}{ccc}
\sum_{\alpha}|a_{\alpha}|^2 &=& t^2+r^2\sin^2\theta,\quad \alpha=0,x,y,z,  \\
\mathrm{Re}\left(a_0^{*}a_k\right) &=& \mathrm{Im}\left(\epsilon_{klm}a_l^{*}a_m\right),\quad k,l,m=x,y,z,  \\
\end{array}
\end{equation}
in which $\epsilon_{klm}$ is the 3D Levi-Civita symbol.
In addition Eq. (\ref{Q_definition_1_SMTQNM}) gives
\begin{equation}\label{Q_definition_4_SMTQNM}
a_0=a_0^{*},\quad a_k=-a_k^{*},\; k=x,y,z.
\end{equation}
Summarizing these two conditions, a reasonable solution to $a_{\alpha}$ is
\begin{equation}\label{Q_definition_5_SMTQNM}
\begin{array}{cc}
a_0 = t\cos\phi_1, & a_x = \mathrm{i}r\sin\theta\sin\phi_2,  \\
a_z = \mathrm{i}t\sin\phi_1, & a_y = \mathrm{i}r\sin\theta\cos\phi_2,  \\
\end{array}
\end{equation}
leading to a physical realization of $Q$ as
\begin{equation}\label{Q_definition_final_SMTQNM}
  Q=\left(
    \begin{matrix}
      t e^{\mathrm{i}\phi_1}             &    r e^{\mathrm{i}\phi_2}\sin\theta \\
      -r e^{-\mathrm{i}\phi_2}\sin\theta &    t e^{-\mathrm{i}\phi_1} \\
    \end{matrix}
  \right).
\end{equation}
Obviously $\phi_{1}$ and $\phi_2$ are the phase shifts
associated with SPT and SMT processes, respectively.
At last, by rotating S-type PSPs 90 degrees clockwise, we get S'-type PSPs and
their scattering matrix can be easily obtained
from Eq. (\ref{Scattering_matrix_S_0_SMTQNM}).

To summarize, in our SMT-QNM at any PSP (S- and S'-type),
for an incoming electron flow with some certain
spin orientation and probability 1, it tunnels into an outgoing flow
with the same spin orientation via SPT process with probability ``$t^2$" 
and also into an outgoing flow with opposite spin orientation via SMT 
process with probability ``$r^2\sin^2\theta$", leaving a probability 
``$r^2\cos^2\theta$" residing in the original equipotential line.

\subsection{II.C 2D Dirac Hamiltonian from SMT-QNM}
The mapping from CC-RNM to 2D Dirac Hamiltonian was accomplished 
in 1996\cite{Ho_Chalker_1996}, and the connection between the 
$Z_2$-QNM and 2D Dirac Hamiltonian was established 
in 2010\cite{Ryu_2010_njp}. 
The main strategy of both works is to view the unitary
(due to probability conservation) scattering matrices as a unitary 
time-evolution operation whose infinitesimal generator is the 
required Hamiltonian, as we all know
that a unitary matrix is the exponential of a Hermitian one.
In this subsection, we follow this strategy and succeed in extracting
the 2D Dirac Hamiltonian from our SMT-QNM and recognizing the roles 
of phase shifts in SMT and SPT at PSPs.
Also, this part of work lays the foundation for understanding the 
close connection between the phase diagram of our SMT-QNM and 
that of disordered 3D weak TIs (see Sec. IV.D). 

\subsubsection{II.C.1 Preparations}
We arrange the S-type and S'-type PSPs alternatively in a 2D Cartesian 
plane to form a bipartite square lattice, as shown in Fig. 2.
Then following the sketch rules in Fig. 1c and 1d, a series of closed 
square plaquettes are obtained, with each edge bearing two 
opposite-directed links. For $r>t$, the centers of these closed plaquettes are 
the potential valleys, while the potential peaks reside in the blanks 
outside. For $r<t$, the situation is just reversed.
Quantum tunnelings (SPT and SMT) occur at the plaquette 
corners, which are the PSPs.
We take the $r>t$ case as the framework for our discussion,
which does not affect the generality of our results.
If one of these plaquettes is assigned with coordinate $(0,0)$,
then the position of anyone in this set is
\begin{equation}\label{Valley_Coordinate_SMTQNM}
\mathbf{R}_{x,y}=x \mathbf{e}_{x} + y \mathbf{e}_{y},\quad x,y\in\mathbb{Z},\quad \mathrm{mod}(x+y,2)=0.
\end{equation}
They form a square lattice and is our main concern.
The eight directed links on the edges of a plaquette are labeled by 
$(n\sigma)$ with $n=1,2,3,4$ and $\sigma=\uparrow$ or $\downarrow$.

\begin{figure}[htbp]
	\centering
	\scalebox{0.65}[0.65]{\includegraphics[angle=0]{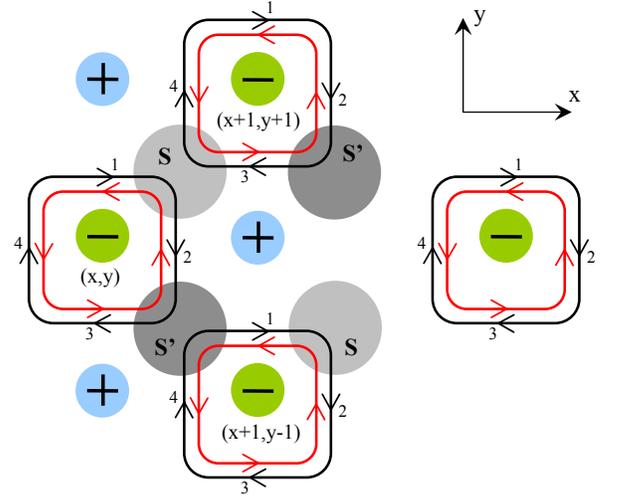}}
	\caption{(Color online) A bipartite square lattice composed of S-type
		and S'-type PSPs. Electron flows illustrated in Fig. 1c and 1d ($r>t$ case) 
		generate a series of corner-shared closed square plaquettes around 
		potential valleys with integral coordinates. The black (red) lines 
	    represent the links for up (down) spin electron flows.
        The scattering-basis order on the edge of each plaquette 
        is specified with Arabic numerals.
        Blue (green) circles indicate potential peaks (valleys).}\label{fig02}
\end{figure}

For the plaquette with coordinate $(x,y)$,
the scattering event at the S-type PSP on its upper-right corner (urc) is
as follows
\begin{equation}\label{Dirac_S_xy_1}
\left(
\begin{matrix}
Z_{2\uparrow}(x,y) \\
Z_{1\downarrow}(x,y) \\
Z_{4\uparrow}(x+1,y+1) \\
Z_{3\downarrow}(x+1,y+1) \\
\end{matrix}
\right)
=S_{x,y}^{\mathrm{urc}}\left(
\begin{matrix}
Z_{1\uparrow}(x,y) \\
Z_{2\downarrow}(x,y) \\
Z_{3\uparrow}(x+1,y+1) \\
Z_{4\downarrow}(x+1,y+1) \\
\end{matrix}
\right),
\end{equation}
with
\begin{equation}\label{Dirac_S_xy_2}
\begin{array}{l}
S_{x,y}^{\mathrm{urc}}\equiv U_{x,y}S_{\mathrm{SMT}}V_{x,y},    \\
U_{x,y} = \mathrm{diag}(e^{\mathrm{i}\psi_2(x,y)},e^{\mathrm{i}\psi_1(x,y)},e^{\mathrm{i}\psi_4(x+1,y+1)},e^{\mathrm{i}\psi_3(x+1,y+1)}), \\
V_{x,y} = \mathrm{diag}(e^{\mathrm{i}\psi_1(x,y)},e^{\mathrm{i}\psi_2(x,y)},e^{\mathrm{i}\psi_3(x+1,y+1)},e^{\mathrm{i}\psi_4(x+1,y+1)}), \\
\end{array}
\end{equation}
in which TRS has be invoked in writing $U_{x,y}$ and $V_{x,y}$.
While that of the S'-type PSP on the lower-right corner (lrc) reads
\begin{equation}\label{Dirac_Sp_xy_1}
\left(
\begin{matrix}
Z_{3\uparrow}(x,y) \\
Z_{2\downarrow}(x,y) \\
Z_{1\uparrow}(x+1,y-1) \\
Z_{4\downarrow}(x+1,y-1) \\
\end{matrix}
\right)
=S_{x,y}^{\mathrm{lrc}}\left(
\begin{matrix}
Z_{2\uparrow}(x,y) \\
Z_{3\downarrow}(x,y) \\
Z_{4\uparrow}(x+1,y-1) \\
Z_{1\downarrow}(x+1,y-1) \\
\end{matrix}
\right),
\end{equation}
with
\begin{equation}\label{Dirac_Sp_xy_2}
\begin{array}{l}
S_{x,y}^{\mathrm{lrc}}\equiv U'_{x,y} S_{\mathrm{SMT}}V'_{x,y}    \\
U'_{x,y} = \mathrm{diag}(e^{\mathrm{i}\psi_3(x,y)},e^{\mathrm{i}\psi_2(x,y)},e^{\mathrm{i}\psi_1(x+1,y-1)},e^{\mathrm{i}\psi_4(x+1,y-1)}) \\
V'_{x,y} = \mathrm{diag}(e^{\mathrm{i}\psi_2(x,y)},e^{\mathrm{i}\psi_3(x,y)},e^{\mathrm{i}\psi_4(x+1,y-1)},e^{\mathrm{i}\psi_1(x+1,y-1)}) \\
\end{array}.
\end{equation}
Next the displacement operators $\tau_{x(y)}^{\pm}$ are introduced as
\begin{equation}\label{Tau_definition}
\begin{array}{ccc}
\tau_x^{\pm}f_{n\sigma}(x,y)  &=&f_{n\sigma}(x\pm 1,y),    \\
\tau_y^{\pm}f_{n\sigma}(x,y)  &=&f_{n\sigma}(x,y\pm 1),    \\
\end{array}
\end{equation}
where $f_{n\sigma}(x,y)$ is an arbitrary function defined at
$\mathbf{R}_{x,y}$.
By definition, they are commutative and
\begin{equation}\label{Tau_features}
\left[\tau_{x(y)}^{\pm}\right]^{-1}= \tau_{x(y)}^{\mp}.
\end{equation}
By rearranging the amplitudes in the order of ``2,4,1,3",
we rewrite Eq. (\ref{Dirac_S_xy_1}) into the form
\begin{equation}\label{Dirac_S_xy_3}
\begin{array}{rcl}
\left(
\begin{matrix}
Z_{2\uparrow}(x,y) \\
Z_{4\uparrow}(x,y) \\
Z_{1\downarrow}(x,y) \\
Z_{3\downarrow}(x,y) \\
\end{matrix}
\right)
&=&\Omega_{S}\left(
\begin{matrix}
Z_{2\downarrow}(x,y) \\
Z_{4\downarrow}(x,y) \\
Z_{1\uparrow}(x,y) \\
Z_{3\uparrow}(x,y) \\
\end{matrix}
\right),    \\
\Omega_{S}&=&O_{x,y}\cdot M_S\cdot O_{x,y},  \\
\end{array}
\end{equation}
with
\begin{equation}\label{Dirac_O}
O_{x,y} = \mathrm{diag}(e^{\mathrm{i}\psi_2(x,y)},e^{\mathrm{i}\psi_4(x,y)},e^{\mathrm{i}\psi_1(x,y)},e^{\mathrm{i}\psi_3(x,y)})
\end{equation}
and
\begin{equation}\label{Dirac_N_S}
\begin{array}{rcl}
M_S&=&\left(
\begin{matrix}
\Phi_2 M_{\mathrm{d}}\Phi_2 & \Phi_1^{\dagger} M_{\mathrm{od}}\Phi_1  \\
\Phi_1 M_{\mathrm{od}} \Phi_1^{\dagger} & -\Phi_2^{\dagger} M_{\mathrm{d}} \Phi_2^{\dagger} \\
\end{matrix}
\right),    \\
\Phi_1&=&\mathrm{diag}(e^{-\mathrm{i}\frac{\phi_1}{2}},e^{\mathrm{i}\frac{\phi_1}{2}}), \quad
\Phi_2=e^{\mathrm{i}\frac{\phi_2}{2}}\sigma_0,   \\
M_{\mathrm{d}}&=&\left(
\begin{matrix}
0 & r\sin\theta\tau_x^+\tau_y^+  \\
-r\sin\theta\tau_x^-\tau_y^- & 0  \\
\end{matrix}
\right),   \\
M_{\mathrm{od}}&=&\left(
\begin{matrix}
r\cos\theta & t\tau_x^+\tau_y^+  \\
t\tau_x^-\tau_y^- & -r\cos\theta  \\
\end{matrix}
\right),   \\
\end{array}
\end{equation}
in which ``d (od)" means diagonal (off-diagonal).
Similarly, Eq. (\ref{Dirac_Sp_xy_1}) is rewritten as
\begin{equation}\label{Dirac_Sp_xy_3}
\begin{array}{rcl}
\left(
\begin{matrix}
Z_{2\downarrow}(x,y) \\
Z_{4\downarrow}(x,y) \\
Z_{1\uparrow}(x,y) \\
Z_{3\uparrow}(x,y) \\
\end{matrix}
\right)
&=&\Omega_{S'}\left(
\begin{matrix}
Z_{2\uparrow}(x,y) \\
Z_{4\uparrow}(x,y) \\
Z_{1\downarrow}(x,y) \\
Z_{3\downarrow}(x,y) \\
\end{matrix}
\right),    \\
\Omega_{S'}&=&O_{x,y}\cdot M_{S'}\cdot O_{x,y},  \\
\end{array}
\end{equation}
where
\begin{equation}\label{Dirac_N_Sp}
\begin{array}{rcl}
M_{S'}&=&\left(
\begin{matrix}
\Phi_2^{\dagger} M'_{\mathrm{d}}\Phi_2^{\dagger} & \Phi_1 M'_{\mathrm{od}}\Phi_1  \\
\Phi_1 \sigma_x M'_{\mathrm{od}} \sigma_x\Phi_1 & -\Phi_2 \sigma_x M'_{\mathrm{d}}\sigma_x\Phi_2 \\
\end{matrix}
\right),    \\
M'_{\mathrm{d}}&=&\left(
\begin{matrix}
0 & -r\sin\theta\tau_x^+\tau_y^-  \\
r\sin\theta\tau_x^-\tau_y^+ & 0  \\
\end{matrix}
\right),   \\
M'_{\mathrm{od}}&=&\left(
\begin{matrix}
t\tau_x^+\tau_y^- & r\cos\theta  \\
-r\cos\theta & t\tau_x^-\tau_y^+  \\
\end{matrix}
\right).   \\
\end{array}
\end{equation}

By defining the total amplitude vector $\mathbf{Z}_{x,y}$ composed of
all eight links along the edges of plaquette at $\mathbf{R}_{x,y}$ as
\begin{equation}\label{Dirac_total_Z}
\mathbf{Z}_{x,y}\equiv \left(Z_{2\uparrow}\; Z_{4\uparrow}\; Z_{1\downarrow}\; Z_{3\downarrow}\; Z_{2\downarrow}\; Z_{4\downarrow}\; Z_{1\uparrow}\; Z_{3\uparrow}\right)^{\mathrm{T}}
\end{equation}
with the superscript ``$\mathrm{T}$" indicating matrix transpose
and introducing $\mu\in\mathbb{Z}$, the elementary imaginary discrete-time evolution
of $\mathbf{Z}_{x,y}$ is,
\begin{equation}\label{Dirac_discrete_time_evolution_1}
\left(\mathbf{Z}_{x,y}\right)_{\mu+1}
=\left(
\begin{matrix}
0 & \Omega_{S}  \\
\Omega_{S'} & 0  \\
\end{matrix}
\right)\left(\mathbf{Z}_{x,y}\right)_{\mu}.
\end{equation}
To acquire decoupled equations, the ``two-step" time evolution,
\begin{equation}\label{Dirac_discrete_time_evolution_2}
\left(\mathbf{Z}_{x,y}\right)_{\mu+2}
=\left(
\begin{matrix}
\Omega_{S} \Omega_{S'} & 0  \\
0 & \Omega_{S'} \Omega_{S} \\
\end{matrix}
\right)\left(\mathbf{Z}_{x,y}\right)_{\mu}
\end{equation}
is more convenient since it is diagonal.
We will focus on $\Omega_{S} \Omega_{S'}\equiv \Omega_{SS'}$ in the rest of this work.
Also we make the transformation
\begin{equation}\label{Dirac_psi_4_redifinition}
\psi_4 \rightarrow \psi_4+\frac{\pi}{2}
\end{equation}
to raise the reference point of the total phase flux of each plaquette
by $\pi$ [see Eqs. (\ref{Dirac_S_xy_3}) and (\ref{Dirac_Sp_xy_3})], 
which is crucial for the extraction of 2D Dirac Hamiltonian.
It can be easily checked that $\Omega_{SS'}$ is unitary, thus 
provide a Hamiltonian as its infinitesimal generator,
\begin{equation}\label{Dirac_Hamiltonian_definition}
\mathcal{H}_{\mathrm{SMT}}=\mathrm{i}\ln \Omega_{SS'}\approx \mathrm{i}\left(\Omega_{SS'}-1\right).
\end{equation}
We then demonstrate that in the close vicinity of the CC-RNM 
critical point
\begin{equation}\label{Dirac_CC_critical_point}
(p_c,\theta_c)_{\mathrm{CC}}=(\frac{1}{2},0),
\end{equation}
how $\mathcal{H}_{\mathrm{SMT}}$ is mapped to 2D Dirac Hamiltonian
by expanding $\Omega_{SS'}$ to the leading-order powers of
\begin{equation}\label{Dirac_CC_expand_variables}
\theta, \; m\equiv 1-\frac{p}{p_c},\; \partial_{x(y)}\equiv \ln\tau_{x(y)}^{+},\; \psi_{1,2,3,4},\; \phi_{1,2}.
\end{equation}

\subsubsection{II.C.2 2D Dirac Hamiltonian around $\theta=0$}
At $\theta=0$, $M_{\mathrm{d}}=M'_{\mathrm{d}}=0$. 
Eqs. (\ref{Dirac_S_xy_3}) and (\ref{Dirac_Sp_xy_3}) thus provide
\begin{equation}\label{Dirac_Omega_SSp_0}
\Omega_{S}^{(0)}=\left(
\begin{matrix}
0 & A^{(0)}  \\
B^{(0)} & 0  \\
\end{matrix}
\right),\quad
\Omega_{S'}^{(0)}=\left(
\begin{matrix}
0 & C^{(0)}  \\
D^{(0)} & 0  \\
\end{matrix}
\right),
\end{equation}
with
\begin{equation}\label{Dirac_A0B0C0D0}
\begin{array}{rcl}
A^{(0)}&=& \left(
\begin{matrix}
e^{\mathrm{i}(\psi_1+\psi_2)}r & e^{\mathrm{i}(\psi_2+\psi_3+\phi_1)} t\tau_x^+ \tau_y^+  \\
\mathrm{i}e^{\mathrm{i}(\psi_1+\psi_4-\phi_1)} t\tau_x^- \tau_y^- & -\mathrm{i}e^{\mathrm{i}(\psi_3+\psi_4)}r  \\
\end{matrix}
\right),        \\
B^{(0)}&=& \left(
\begin{matrix}
e^{\mathrm{i}(\psi_1+\psi_2)}r & \mathrm{i}e^{\mathrm{i}(\psi_1+\psi_4-\phi_1)} t\tau_x^+ \tau_y^+  \\
e^{\mathrm{i}(\psi_2+\psi_3+\phi_1)} t\tau_x^- \tau_y^- & -\mathrm{i}e^{\mathrm{i}(\psi_3+\psi_4)}r  \\
\end{matrix}
\right),        \\
C^{(0)}&=& \left(
\begin{matrix}
e^{\mathrm{i}(\psi_1+\psi_2-\phi_1)} t\tau_x^+ \tau_y^- & e^{\mathrm{i}(\psi_2+\psi_3)}r  \\
-\mathrm{i}e^{\mathrm{i}(\psi_1+\psi_4)}r & \mathrm{i}e^{\mathrm{i}(\psi_3+\psi_4+\phi_1)} t\tau_x^- \tau_y^+  \\
\end{matrix}
\right),        \\
D^{(0)}&=& \left(
\begin{matrix}
e^{\mathrm{i}(\psi_1+\psi_2-\phi_1)} t\tau_x^- \tau_y^+ & -\mathrm{i}e^{\mathrm{i}(\psi_1+\psi_4)}r  \\
e^{\mathrm{i}(\psi_2+\psi_3)}r & \mathrm{i}e^{\mathrm{i}(\psi_3+\psi_4+\phi_1)} t\tau_x^+ \tau_y^-  \\
\end{matrix}
\right).        \\
\end{array}
\end{equation}
Then
\begin{equation}\label{Dirac_HSMT_0}
\mathcal{H}_{\mathrm{SMT}}^{(0)}=\mathrm{i}\left(
\begin{matrix}
A^{(0)}D^{(0)}-\sigma_0 & 0  \\
0 & B^{(0)}C^{(0)}-\sigma_0  \\
\end{matrix}
\right)\equiv
\left(
\begin{matrix}
J_+ & 0  \\
0 & J_-  \\
\end{matrix}
\right).
\end{equation}
Under the following assumptions:

\noindent
(a) displacement operators act on smooth enough functions thus
\begin{equation}\label{Dirac_assumptions_1}
\tau_{x(y)}^{\pm}\rightarrow 1\pm\partial_{x(y)},
\end{equation}

\noindent
(b) the phases $\psi_{n=1,2,3,4}$ and $\phi_1$ are small enough hence
\begin{equation}\label{Dirac_assumptions_2}
e^{\pm\mathrm{i}\psi_n}\rightarrow 1\pm \mathrm{i}\psi_n, \quad e^{\pm\mathrm{i}\phi_1}\rightarrow 1\pm \mathrm{i}\phi_1,
\end{equation}

\noindent
(c) in the close vicinity of CC-RNM critical point one has
\begin{equation}\label{Dirac_assumptions_3}
r\rightarrow \frac{1}{\sqrt{2}}\left(1-\frac{m}{2}\right),\quad t\rightarrow \frac{1}{\sqrt{2}}\left(1+\frac{m}{2}\right),
\end{equation}
we get
\begin{equation}\label{Dirac_J_plus_minus}
\begin{array}{l}
J_+ =A_0\sigma_0+(-\mathrm{i}\partial_x + A_x)\sigma_y - (-\mathrm{i}\partial_y + A_y)\sigma_z - m\sigma_x,     \\
J_- =A_0\sigma_0-(-\mathrm{i}\partial_x - A_x)\sigma_z + (-\mathrm{i}\partial_y - A_y)\sigma_x + m\sigma_y,     \\
\end{array}
\end{equation}
with
\begin{equation}\label{Dirac_scalar_vector_potentials}
\begin{array}{rcl}
A_0      &=&-(\psi_1+\psi_2+\psi_3+\psi_4),     \\
(A_x,A_y)&=&(-\psi_1+\psi_3+\phi_1,\psi_2-\psi_4),   \\
\end{array}
\end{equation}
acting as a scalar/vector potential, respectively.

Then the system is driven away slightly from the critical point
(\ref{Dirac_CC_critical_point}) along the $\theta$-line.
Hence $\psi_n=m=0$ and $\tau_{x(y)}^{\pm}=1$, and to the leading order of $\theta$ one has
\begin{equation}\label{Dirac_HSMT_1}
\Omega_{SS'}=\Omega_{SS'}^{(0)}+\theta\left(\Omega_{S}^{(1)}\Omega_{S'}^{(0)}+\Omega_{S}^{(0)}\Omega_{S'}^{(1)}\right)+\cdots,
\end{equation}
with
\begin{equation}\label{Dirac_Omega_S_Sp_1}
\begin{array}{rcl}
\Omega_{S}^{(1)}&=&\frac{1}{\sqrt{2}}\left(
\begin{matrix}
-e^{\mathrm{i}\phi_2} & 0  \\
0 & -\mathrm{i}e^{-\mathrm{i}\phi_2}  \\
\end{matrix}
\right)\otimes \sigma_y,        \\
\Omega_{S'}^{(1)}&=&\frac{1}{\sqrt{2}}\left(
\begin{matrix}
e^{-\mathrm{i}\phi_2} & 0  \\
0 & -\mathrm{i}e^{\mathrm{i}\phi_2}  \\
\end{matrix}
\right)\otimes \sigma_y.         \\
\end{array}
\end{equation}
Correspondingly, the SMT Hamiltonian turns to
\begin{equation}\label{Dirac_HSMT_01}
\mathcal{H}_{\mathrm{SMT}}=\left(
\begin{matrix}
J_+ & J_{\theta}  \\
J_{\theta}^{\dagger} & J_-  \\
\end{matrix}
\right), \quad
J_{\theta}\equiv\theta e^{\mathrm{i}\phi_2}\left(
\begin{matrix}
\mathrm{i} & -\mathrm{i}  \\
1 & 1  \\
\end{matrix}
\right).
\end{equation}
After performing a unitary transformation
\begin{equation}\label{Dirac_U_final}
\begin{array}{rcl}
\mathcal{U}&=&\left(
\begin{matrix}
e^{\mathrm{i}\frac{\pi}{4}\sigma_x} & 0  \\
0 & e^{-\mathrm{i}\frac{\pi}{4}\sigma_y}  \\
\end{matrix}
\right)\cdot
\left(
\begin{matrix}
e^{\mathrm{i}\frac{\pi}{4}\sigma_y} & 0  \\
0 & e^{-\mathrm{i}\frac{\pi}{4}\sigma_x}  \\
\end{matrix}
\right)      \\
  &  & \; \; \cdot
  \left(
  \begin{matrix}
  e^{-\frac{\mathrm{i}}{2}\left(\phi_2+\frac{5\pi}{4}\right)}\sigma_0 & 0  \\
  0 & e^{ \frac{\mathrm{i}}{2}\left(\phi_2+\frac{5\pi}{4}\right)}\sigma_0  \\
  \end{matrix}
  \right),      \\
\end{array}
\end{equation}
we get the final Hamiltonian
\begin{equation}\label{Dirac_H_final}
\mathcal{H}_{\mathrm{f}}
=\mathcal{U}^{\dagger}\mathcal{H}_{\mathrm{SMT}}\mathcal{U}
=\left(
\begin{matrix}
\mathcal{H}_{+}^{\mathrm{D}} & \sqrt{2}\theta\sigma_0  \\
\sqrt{2}\theta\sigma_0 & \mathcal{H}_{-}^{\mathrm{D}}  \\
\end{matrix}
\right),
\end{equation}
with
\begin{equation}\label{Dirac_H_Dirac_plus_minus}
\mathcal{H}_{\pm}^{\mathrm{D}} =A_0\sigma_0+(-\mathrm{i}\partial_x \pm A_x)\sigma_x + (-\mathrm{i}\partial_y \pm A_y)\sigma_y \pm m\sigma_z.
\end{equation}
Obviously $\mathcal{H}_{\mathrm{f}}$ describes a pair of
Dirac fermions (with mass $\pm m$) subject to the same random scalar
potential $A_0$ and respective random vector potential $\pm (A_x,A_y)$,
meantime bearing a mutual coupling $\sqrt{2}\theta\sigma_0$.
By introducing a ``valley" space distinguishing these two Dirac 
fermions (different locations of Dirac cones in Brillouin zone),
the final Hamiltonian can be rewritten as
\begin{equation}\label{Dirac_H_final_valley_description}
\begin{array}{lll}
\mathcal{H}_{\mathrm{f}}&=& s_0\otimes(-\mathrm{i}\partial_x\sigma_x-\mathrm{i}\partial_y\sigma_y)+s_0\otimes A_0\sigma_0   \\
& &+s_z\otimes(A_x\sigma_x+A_y\sigma_y+m\sigma_z)+s_x\otimes\sqrt{2}\theta\sigma_0. \\
\end{array}
\end{equation}
where $s_0$ and $s_{x,y,z}$ are identity and
Pauli matrices in valley space.
Therefore our SMT-QNM belongs to the symplectic class and should
be an effective model for ATs in QSH ensembles.
Also, the above analytics shows that the phase shifts in SPT and
SMT processes at PSPs have different roles during the extraction of 
2D Dirac Hamiltonian. The former ($\phi_1$) enters the vector 
potentials thus could have impacts on geometric phase 
accumulated along the plaquette edges. While the latter ($\phi_2$) 
resides in the coupling matrix $J_{\theta}$ between $J_{\pm}$ 
and then manifests itself in the unitary transformation that changes 
$\mathcal{H}_{\mathrm{SMT}}$ to $\mathcal{H}_{\mathrm{f}}$,
hence acts as a gauge field describing the spin-flip interaction.

\section{III. Algorithms for finite-size analysis}
\subsection{III.A Two-terminal conductance $G_{\mathrm{2T}}$}

\begin{figure}[htbp]
	\centering
	\scalebox{0.5}[0.5]{\includegraphics[angle=0]{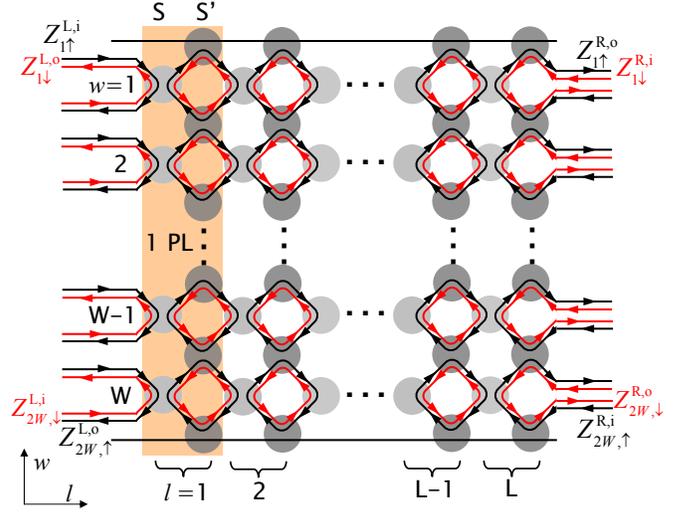}}
	\caption{(Color online) An example of the layout of a SMT-QNM 
		network with boundary nodes being S'-type PSPs. 
		The light (dark) gray circles indicate S-type (S'-type) PSPs. 
		This 2D PSP lattice is composed of $L$ PLs. Each PL (orange
		vertical strip) consists of $W$ S-type and $W$ S'-type PSPs. 
		$4W$ electron flows ($2W$ incoming and $2W$ outgoing) distribute 
		regularly on each side of the network and are related by 
		the total transfer matrix $T_W^L$.}\label{fig03}
\end{figure}

For numerical convenience, by rotating Fig.2 45 degrees clockwise, we obtain a
2D PSP lattice composed of $L$ principal layers (PLs), as shown in Fig. 3.
Each PL consists of $W$ S-type and $W$ S'-type PSPs.
At a S-type PSP, the ``left-ro-right" transfer matrix is obtained from
its scattering matrix [see Eq. (\ref{Scattering_matrix_S_0_SMTQNM})] as
\begin{equation}\label{Transfer_matrix_T_S}
\left(
  \begin{matrix}
    Z_{4\uparrow}^{\mathrm{o}} \\
    Z_{4\downarrow}^{\mathrm{i}} \\
    Z_{3\uparrow}^{\mathrm{i}} \\
    Z_{3\downarrow}^{\mathrm{o}} \\
  \end{matrix}
\right)
=T_{S}\left(
  \begin{matrix}
    Z_{1\uparrow}^{\mathrm{i}} \\
    Z_{1\downarrow}^{\mathrm{o}} \\
    Z_{2\uparrow}^{\mathrm{o}} \\
    Z_{2\downarrow}^{\mathrm{i}} \\
  \end{matrix}
\right),  \quad T_{S}=\mathcal{U}_1\cdot T_0 \cdot \mathcal{U}_2,   \\
\end{equation}
in which
\begin{equation}\label{Transfer_matrix_U12}
\begin{array}{c}
\mathcal{U}_1=\mathrm{diag}(e^{-\mathrm{i}\frac{\phi_1-\phi_2}{2}},e^{\mathrm{i}\frac{\phi_1-\phi_2}{2}},e^{-\mathrm{i}\frac{\phi_1-\phi_2}{2}},e^{\mathrm{i}\frac{\phi_1-\phi_2}{2}}),    \\
\mathcal{U}_2=\mathrm{diag}(e^{-\mathrm{i}\frac{\phi_1+\phi_2}{2}},e^{\mathrm{i}\frac{\phi_1+\phi_2}{2}},e^{-\mathrm{i}\frac{\phi_1+\phi_2}{2}},e^{\mathrm{i}\frac{\phi_1+\phi_2}{2}}),    \\
\end{array}
\end{equation}
and
\begin{equation}\label{Transfer_matrix_T_0}
\begin{array}{rcl}
T_0&=&\frac{1}{1-r^2\cos^2\theta}\left[\sigma_0\otimes (T_0)^{\mathrm{d}}+\sigma_x\otimes (T_0)^{\mathrm{od}}\right],    \\
(T_0)^{\mathrm{d}}&=&\left(
\begin{matrix}
t & r^2\sin\theta\cos\theta  \\
-r^2\sin\theta\cos\theta & t  \\
\end{matrix}
\right),   \\
(T_0)^{\mathrm{od}}&=&\left(
\begin{matrix}
-rt\cos\theta & -r\sin\theta  \\
r\sin\theta & -rt\cos\theta  \\
\end{matrix}
\right).   \\
\end{array}
\end{equation}
While at a S'-type PSP, the counterpart is
\begin{equation}\label{Transfer_matrix_T_Sp}
\left(
  \begin{matrix}
    Z_{4\uparrow}^{\mathrm{i}} \\
    Z_{4\downarrow}^{\mathrm{o}} \\
    Z_{3\uparrow}^{\mathrm{o}} \\
    Z_{3\downarrow}^{\mathrm{i}} \\
  \end{matrix}
\right)
=T_{S'}
\left(
  \begin{matrix}
    Z_{1\uparrow}^{\mathrm{o}} \\
    Z_{1\downarrow}^{\mathrm{i}} \\
    Z_{2\uparrow}^{\mathrm{i}} \\
    Z_{2\downarrow}^{\mathrm{o}} \\
  \end{matrix}
\right),  \quad T_{S'}=\mathcal{U}'_1\cdot T'_0 \cdot (\mathcal{U}'_1)^{\dagger},   \\
\end{equation}
where
\begin{equation}\label{Transfer_matrix_Up1}
\mathcal{U}'_1=\mathrm{diag}(e^{\mathrm{i}\frac{\phi_1+\phi_2}{2}},e^{-\mathrm{i}\frac{\phi_1+\phi_2}{2}},e^{-\mathrm{i}\frac{\phi_1-\phi_2}{2}},e^{\mathrm{i}\frac{\phi_1-\phi_2}{2}})
\end{equation}
and
\begin{equation}\label{Transfer_matrix_Tp_0}
\begin{array}{rcl}
T'_0&=&\frac{1}{r}\left[\sigma_z\otimes (T'_0)^{\mathrm{d}}+\mathrm{i}\sigma_y\otimes (T'_0)^{\mathrm{od}}\right],    \\
(T'_0)^{\mathrm{d}}&=&\left(
\begin{matrix}
\cos\theta & -t\sin\theta  \\
t\sin\theta & \cos\theta  \\
\end{matrix}
\right),   \\
(T'_0)^{\mathrm{od}}&=&\left(
\begin{matrix}
-t\cos\theta & \sin\theta  \\
-\sin\theta & -t\cos\theta  \\
\end{matrix}
\right).   \\
\end{array}
\end{equation}
Then the transfer matrix for the $k-$th PL is
\begin{equation}\label{k-th_PL_Tk}
T^{(k)}=V_{4}^{(k)}V_{3}V_{2}^{(k)}V_{1},
\end{equation}
where the boundary nodes are selected to be S'-type PSPs as an 
example (see Fig. 3).
$V_1$ is the transfer matrix of the sub-layer composed merely by 
S-type PSPs with the following form
\begin{equation}\label{k-th_PL_V1}
V_1=\mathrm{diag}(\,\underbrace{T_{0},\cdots,T_{0}}_{W}\,),
\end{equation}
$V_3$ is the transfer matrix of the S'-type sub-layer
\begin{equation}\label{k-th_PL_V3}
V_3=
\begin{array}{cc}
\left(
\begin{array}{ccccc}
B_1      & 0               & \cdots & 0               & B_2    \\
0        & T'_{0}          & \cdots & 0               & 0      \\
\vdots   & \vdots          & \ddots & \vdots          & \vdots \\
0        & 0               & \cdots & T'_{0}          & 0      \\
B_3      & 0               & \cdots & 0               & B_4   \\
\end{array}
\right)
\left.
\begin{array}{c}
\!\!\!\!\!\!\!\!\\[8mm]
\end{array}
\right\}{W-1}
\end{array},
\end{equation}
where $B_{1,2,3,4}$ are $2\times 2$ matrices and determined 
by the choice of boundary condition in transverse direction.
When we focus on edge modes, the reflecting boundary 
condition (RBC) is imposed. The Kramers pair is totally 
reflected without any spin flip at boundary nodes, thus
\begin{equation}\label{RBC}
\left(
\begin{matrix}
B_4   &  B_3    \\
B_2   &  B_1    \\
\end{matrix}
\right)=\sigma_0\otimes\sigma_0.
\end{equation}
If bulk behaviors are the main concern, the periodic boundary 
condition (PBC) is adopted, which means
\begin{equation}\label{PBC}
\left(
\begin{matrix}
B_4   &  B_3    \\
B_2   &  B_1    \\
\end{matrix}
\right)=T'_{0}.
\end{equation}
At last, $V_2^{(k)}$ and $V_4^{(k)}$ are $4W\times 4W$ diagonal 
matrices,
\begin{equation}\label{Random_phase_matrice_1}
\left[V_{\alpha}^{(k)}\right]_{lm}=\delta_{lm}e^{\mathrm{i}\psi_{\alpha,l}^{(k)}},\quad (\alpha=2,4),
\end{equation}
describing the \emph{left-to-right} intra- and inter-PL random phases in $4W$
links connecting S-type and S'-type PSPs in adjacent sub-layers.
Note that TRS ensures in any link,
spin-up electron flowing in a certain direction acquires the same
dynamical phase with that of a spin-down electron in the opposite direction.
Thus one has the ``phase pairing rule"
\begin{equation}\label{Random_phase_matrice_2}
\psi_{\alpha,2w-1}^{(k)}+\psi_{\alpha,2w}^{(k)}=0,\quad (w=1,\ldots,2W).
\end{equation}
In practice, for certain $\alpha$ the $2W$ phases $\phi_{\alpha,2w-1}^{(k)}$ 
are independently and uniformly distributed in $[0,2\pi)$.

Multiplying $T^{(k)}$ sequentially, the total transfer matrix $T_W^L$, 
which relates the electron flows on the left of the network
$(Z_{1\uparrow}^{\mathrm{L,i}},Z_{1\downarrow}^{\mathrm{L,o}},\cdots,Z_{2W\uparrow}^{\mathrm{L,o}},Z_{2W\downarrow}^{\mathrm{L,i}})^{\mathrm{T}}$
and those on the right
$(Z_{1\uparrow}^{\mathrm{R,o}},Z_{1\downarrow}^{\mathrm{R,i}},\cdots,Z_{2W\uparrow}^{\mathrm{R,i}},Z_{2W\downarrow}^{\mathrm{R,o}})^{\mathrm{T}}$, is then obtained
\begin{equation}\label{Total_transfer_matrix}
\left(
\begin{matrix}
Z_{1\uparrow}^{\mathrm{R,o}} \\
Z_{1\downarrow}^{\mathrm{R,i}} \\
\vdots \\
Z_{2W\uparrow}^{\mathrm{R,i}} \\
Z_{2W\downarrow}^{\mathrm{R,o}} \\
\end{matrix}
\right)
=T_W^L
\left(
\begin{matrix}
Z_{1\uparrow}^{\mathrm{L,i}} \\
Z_{1\downarrow}^{\mathrm{L,o}} \\
\vdots \\
Z_{2W\uparrow}^{\mathrm{L,o}} \\
Z_{2W\downarrow}^{\mathrm{L,i}} \\
\end{matrix}
\right),\quad T_W^L=T^{(L)}\cdots T^{(1)}.
\end{equation}
By introducing a unitary matrix $O$ with
\begin{equation}\label{O_definition}
O_{mn}=\left\{
\begin{array}{cl}
1, & (m,n)=\left\{
           \begin{array}{lc}
	             (4w-3,w)    &\mathrm{or} \\
	             (4w-2,3W+w) &\mathrm{or} \\
	             (4w-1,2W+w) &\mathrm{or} \\
	             (4w,W+w)    &  \; \\
	             \;\; w=1,\cdots,W & \;\\
	       \end{array} 
	       \right.   \\ 
0,  &  \mathrm{otherwise}    \\
\end{array} 
\right. ,
\end{equation}
the electron flows on each side of the system are reordered into 
four subgroups marked by 
$(\alpha=\mathrm{i/o},\sigma=\uparrow\downarrow)$, i.e.,
\begin{equation}\label{Transimission_0}
\left(
\begin{matrix}
Z_{\mathrm{o},\uparrow}^{\mathrm{R}} \\
Z_{\mathrm{o},\downarrow}^{\mathrm{R}} \\
Z_{\mathrm{i},\uparrow}^{\mathrm{R}} \\
Z_{\mathrm{i},\downarrow}^{\mathrm{R}} \\
\end{matrix}
\right)
=\widetilde{T}
\left(
\begin{matrix}
Z_{\mathrm{i},\uparrow}^{\mathrm{L}} \\
Z_{\mathrm{i},\downarrow}^{\mathrm{L}} \\
Z_{\mathrm{o},\uparrow}^{\mathrm{L}} \\
Z_{\mathrm{o},\downarrow}^{\mathrm{L}} \\
\end{matrix}
\right),\quad \widetilde{T}=O^{\dagger}T_W^L O.
\end{equation}
On the other hand, the entire network can be viewed as a whole hence 
its transport features are provided by a $4W\times4W$ scattering 
matrix $S_\mathrm{t}$,
\begin{equation}\label{Transimission_1}
\left(
\begin{matrix}
Z_{\mathrm{o},\uparrow}^{\mathrm{L}} \\
Z_{\mathrm{o},\downarrow}^{\mathrm{L}} \\
Z_{\mathrm{o},\uparrow}^{\mathrm{R}} \\
Z_{\mathrm{o},\downarrow}^{\mathrm{R}} \\
\end{matrix}
\right)
=S_{\mathrm{t}}
\left(
\begin{matrix}
Z_{\mathrm{i},\uparrow}^{\mathrm{L}} \\
Z_{\mathrm{i},\downarrow}^{\mathrm{L}} \\
Z_{\mathrm{i},\uparrow}^{\mathrm{R}} \\
Z_{\mathrm{i},\downarrow}^{\mathrm{R}} \\
\end{matrix}
\right),\quad
S_{\mathrm{t}}=
\left(
\begin{matrix}
R & T' \\
T & R' \\
\end{matrix}
\right),
\end{equation}
where $T$ and $T'$ ($R$ and $R'$) are $2W\times2W$ transmission (reflection) matrices.
The Landauer formula tells us that the total two-terminal charge conductance, 
$G_{\mathrm{2T}}$, is
\begin{equation}\label{Transimission_2}
G_{\mathrm{2T}}=\frac{e^2}{h}\mathrm{Tr}(T'^{\dagger}T').
\end{equation}
Finally, by comparing Eqs. (\ref{Transimission_0}) and (\ref{Transimission_1}),
one has
\begin{equation}\label{Transimission_3}
\widetilde{T}=
\left(
\begin{matrix}
\widetilde{T}_{11} & \widetilde{T}_{12} \\
\widetilde{T}_{21} & \widetilde{T}_{22} \\
\end{matrix}
\right)
=\left(
\begin{matrix}
T-R'T'^{-1}R & R'T'^{-1} \\
-T'^{-1}R    & T'^{-1} \\
\end{matrix}
\right),
\end{equation}
which leads to $T'=(\widetilde{T}_{22})^{-1}$.

This provides the main algorithm of calculating the two-terminal conductance.
Before ending this subsection, a few points need to be addressed.
First, the diagonal phases in $\mathcal{U}_{1,2}$ and $\mathcal{U}'_{1}$
are already grouped to pairs with opposite signs, thus can be absorbed into
random phase matrices $V_{2,4}^{(k)}$.
This feature has two consequences: (i) we directly use $T_0 (T'_0)$
rather than $T_S (T_{S'})$ to build $V_1(V_3)$,
(ii) in real calculations, usually $\phi_{1,2}$ are assumed to be 
distributed independently and uniformly in $[0,2\pi)$ or even neglected.
Second, during the calculation of $\widetilde{T}_{22}$, the numerical instability
of multiplying iteratively $T^{(k)},k=1,\cdots,L$ can be fixed by performing QR
decompositions where needed. 

\subsection{III.B Lyapunov exponents and normalized localization length}
For a quasi-one-dimensional (Q1D) system ($W$ finite, $L\rightarrow\infty$),
generally the Anderson localization effect makes the two-terminal transmission
decays exponentially. The corresponding decay length is called the Q1D
localization length $\xi_{W}$, which is the function of Fermi level ($p$),
SMT ($\theta$) and transverse dimension $W$.

Now we define a real $4W\times4W$ symmetric matrix
\begin{equation}\label{Lyapunov_exponent_0}
\Xi=\mathrm{ln}\left[\left(T_W^L\right)^{\dagger}T_W^L\right].
\end{equation}
The TRS makes the $4W$ eigenvalues of $\Xi$ doubly degenerate into $2W$ pairs,
and further fall into $W$ groups with opposite signs due to the current conservation
request. In other words, the eigenvalues of $\Xi$ can be written as
$\pm \omega_i,i=1,\ldots,2W$ meantime satisfying $0<\omega_1=\omega_2<\omega_3=\omega_4<\ldots<\omega_{2W-1}=\omega_{2W}$.
The Lyapunov exponents (LEs) associated with this Q1D network system with fixed 
width $W$ are then defined by the following limit
\begin{equation}\label{Lyapunov_exponent_1}
\Gamma_i=\lim_{L\rightarrow\infty}\omega_i/(2L),
\end{equation}
and are self-averaging random variables.

The Q1D localization length of electrons is defined as the reciprocal of the 
\emph{smallest positive} LE,
\begin{equation}\label{LL_definition}
\xi_W=1/\Gamma_1,
\end{equation}
since the decay of transmission should be controlled by the lowest decay 
rate in this system. Finally, the criticality of the 2D system is determined 
by the behavior of normalized localization length $\Lambda$,
\begin{equation}\label{Normalized_LL_definition}
\Lambda\equiv \xi_W/W,
\end{equation}
as the transverse dimension $W$ increases:
the system falls into NM (NI) phase when $\Lambda$ is an 
increasing (decreasing) function of $W$ for sufficient large $W$.

In practical numerical calculations, the LEs are not obtained by 
directly diagonalizing $\Xi$, which comes from iterative multiplication 
of transfer matrices and turns to be numerically unstable.
Following Ref.\cite{Slevin_2014_njp}, we employ the following algorithm to 
achieve satisfactory estimations for both the LEs and their precision.
For simplicity, suppose $L=s\cdot r\cdot m$, where $s,r,m$ are integers.
To estimate all $2W$ LEs, a $4W\times 2W$ matrix
$K^{(0)}$ with random orthogonal columns is multiplied to $T^{(1)}$.
We then perform the following QR decomposition every $m$ steps,
\begin{equation}\label{LE_algorithm_1}
K^{(j)}M^{(j)}=\left[T^{(jm)}\cdots T^{((j-1)m+1)}\right]K^{(j-1)},
\end{equation}
where $j=1,\ldots,sr$, $K^{(j)}$ are $4W\times 2W$ matrices with orthogonal 
columns and $M^{(j)}$ are $2W\times 2W$ upper triangular matrices with 
positive diagonal elements.

The total length $L$ is divided into $s$ segments and each consists of 
$r\cdot m$ PLs. In each $k-$segment ($1\le k \le s$), we calculate,
\begin{equation}\label{LE_algorithm_2}
\gamma_{2W+1-w}^{(k)}=\frac{1}{rm}\sum_{j=(k-1)r+1}^{kr}\mathrm{ln}M_{w,w}^{(j)},\; 1\le w\le 2W.
\end{equation}
The $2W$ LEs are then evaluated as,
\begin{equation}\label{LE_algorithm_3}
\Gamma_i=\overline{\gamma_i}=\frac{1}{s}\sum_{k=1}^{s}\gamma_i^{(k)},\quad 1\le i\le 2W.
\end{equation}
If each segment ($rm$) is long enough, it is reasonable to assume that 
$\gamma_i^{(k)}$($1\le k \le s$) are
statistically independent. The standard error $\sigma_i$ of $\Gamma_i$ 
is given by,
\begin{equation}\label{LE_algorithm_4}
\sigma_i=\frac{1}{\sqrt{s-1}}\left(\overline{\gamma_i^2}-\overline{\gamma_i}^2\right)^{\frac{1}{2}},\quad \overline{\gamma_i^2}=\frac{1}{s}\sum_{k=1}^{s}\left(\gamma_i^{(k)}\right)^2.
\end{equation}
In most cases, $\epsilon_1=\sigma_1/\Gamma_1=1\%$ is an acceptable criterion
for a good estimation of $\Gamma_1$ and thus $\Lambda$.

\section{IV. Quantum phases and phase transitions in SMT-QNM}
The simplest S-type PSP is realized by 2D quadratic potential
$V_{\mathrm{S-PSP}}=U_0\cdot (y^2-x^2)$ with $U_0>0$,
which is identical for arbitrary spin orientation.
In CC-RNM, the total Hamiltonian of an electron close to a S-type PSP
\begin{equation}\label{H_S_PSP}
\mathcal{H}_{\mathrm{S-PSP}}=\left(-\mathrm{i}\hbar\nabla+e\mathbf{A}\right)^2/(2m)+V_{\mathrm{S-PSP}},
\end{equation}
is quadratic hence can be diagonalized.
Under symmetric gauge of the vector potential 
$\mathbf{A}=(B/2)(-y,x)$, the reflecting probability 
is\cite{Kramer_2005,Fertig_1987}
\begin{equation}\label{p_expression_1}
p\equiv r^2=\left(1+e^{\pi\varepsilon}\right)^{-1},
\end{equation}
with
\begin{equation}\label{p_expression_2}
\begin{array}{ccl}
\varepsilon &=&[E_F-(2n+1)E_{+}]/E_-,     \\
E_{\pm}&=&\sqrt{\pm\left[\left(\frac{\hbar\omega_B}{4}\pm\hbar\omega'\right)^2-\lambda^2\right]},\quad \lambda=\frac{\hbar U_0}{eB},   \\
\hbar\omega' &=&\sqrt{\lambda^2+\left(\frac{\hbar\omega_B}{4}\right)^2},\quad \omega_B=\frac{eB}{m},
\end{array}
\end{equation}
in which $E_F$ is the Fermi energy of the system.
If $E_F$ is well below (above) the saddle point energy, the quantum tunneling 
probability vanishes (approaches to 1). 
When $E_F$ lies exactly at the PSP energy,
$\varepsilon=0$ hence $p_c=1/2$ being the CC-RNM critical point.

In SMT-QNM, there are no external magnetic fields. However we preserve the
mathematical structure in Eq. (\ref{p_expression_1}) and in the simplest case
assume $\varepsilon\equiv E_F$ without loss of generality.
Then the mapping from $E_F\in[-\infty,+\infty]$ to $p\in[0,1]$ is bijective
with $E_F=0$ corresponding to $p=p_c$. Hence the ``SMT-SPT" phase space is 
isomorphic to ``$\theta-p$" parameter space.

To get the full phase diagram, both bulk and edge behaviors are important.
In the first step, PBC is exerted to distinguish insulating (NI and TI) and 
NM states. Then RBC is adopted to further check whether
there are topological non-trivial edge modes.
The network layout is depicted in Fig. 3, with S'-type PSPs being the 
boundary nodes. There is also counterpart with marginal S-type PSPs.
However they are equivalent under PBC while differ only in 
boundary modes under RBC (symmetric about $p=p_c$). 
Throughout the rest of this work, we fix the boundary nodes to be S'-type PSPs
under RBC.
In this section, the quantum phases and phase transitions in the 
closed phase space 
\begin{equation}\label{Phase_space_p_theta}
\Omega_1\equiv \left\{(p,\theta)|0\le p\le1,0\le\theta\le\pi/2\right\}
\end{equation}
are investigated in details.

\begin{figure*}[htbp]
	\centering
	\scalebox{1}[1]{\includegraphics[angle=0]{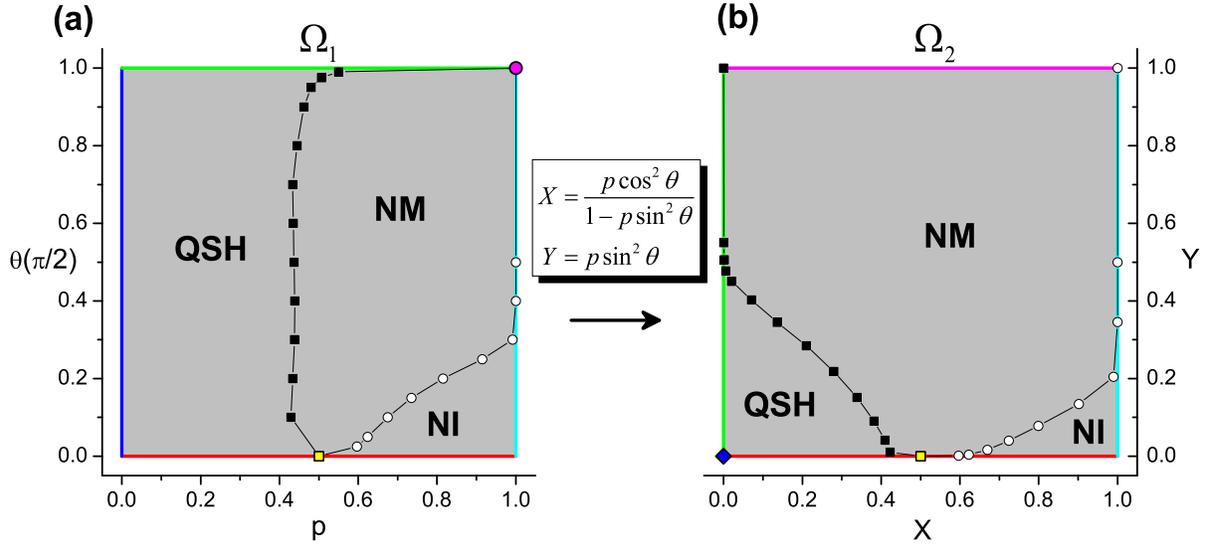}}
	\caption{(Color online) Phase diagrams of the SMT-QNM in phase 
		space $\Omega_1$ (a) and $\Omega_2$ (b). 
		In each of them, NM phase is sandwiched by QSH and NI phases.
		The symmetries on the boundaries and in the interior of $\Omega_1$ are 
		discussed in the main context in details. 
		Through the mapping (\ref{p_theta_to_X_Y}),
		the asymmetric phase diagram in $\Omega_1$ becomes the symmetric counterpart
		in $\Omega_2$. Lines with the same colors indicate
		the correspondence between the boundaries of the two phase spaces.
		The CC-RNM critical point $(p_c,\theta_c)=(X_c,Y_c)=(1/2,0)$ is denoted
		by yellow solid squares in both phase spaces.}\label{fig04}
\end{figure*}

\subsection{IV.A Phase diagram of SMT-QNM in $\Omega_1$}
Following the algorithms in Sec. III, $G_{\mathrm{2T}}$ and $\Lambda$ are 
calculated under PBC and/or RBC. 
Based on these numerical data, the complete phase diagram of SMT-QNM is 
obtained, as plotted in Fig. 4a.
Several important features are collected and explained as follows.

\subsubsection{IV.A.1 Symmetry about $p=p_c$ when $\theta=0$}
When SMT is absent ($\theta=0$), the SMT-QNM is nothing but two decoupled 
copies of CC-RNM with opposite chiralities meantime bearing opposite 
spin orientations.
At all PSPs, When $p\rightarrow 0$ the quantum tunneling $t=\sqrt{1-p}$ 
defeats the reflecting amplitude $r=\sqrt{p}$ along equipotential lines.
Hence all electron current loops around potential peaks become closed.
On the contrary, when $p\rightarrow 1$ at PSPs the quantum tunneling gets 
weak and the reflecting along equipotential lines dominates.
All electron current loops around potential valleys then become closed.
Under PBC, these two cases are equivalent and both lead to NI phase.
Between these two phases, $p_c=0.5$ ($P_{\mathrm{CC}}$ point in Fig. 4) 
is the quantum critical 
point, which can be obtained from the infinitesimal Migdal-Kadanoff 
transformation for real-space renormalization of CC-RNM\cite{Arovas_1997}.
While under RBC, different choices of marginal PSP nodes result in
different boundary modes on network edges.
In Fig. 5 we illustrate the case in which S'-type PSPs reside
in boundaries thus a quantum doublet emerges on each edge leading to 
the QSH state when $p<p_c$.

\begin{figure}[htbp]
	\centering
	\scalebox{0.55}[0.55]{\includegraphics[angle=0]{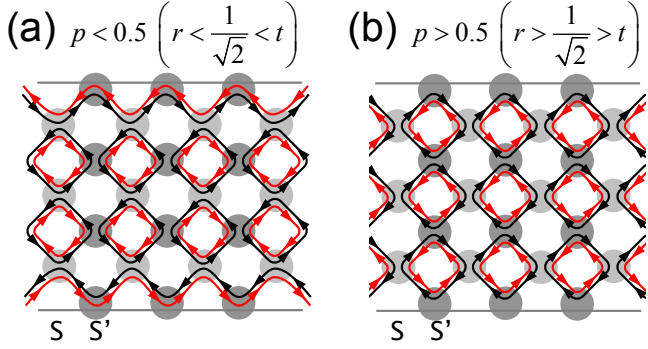}}
	\caption{(Color online) Electron-flow distribution of SMT-QNM under RBC 
		in the absence of SMT ($\theta=0$) with boundary nodes being S'-type PSPs. 
		(a) The QSH state when $p<p_c$. A Kramers doublet resides on each 
		edge of the sample. (b) The NI phase when $p>p_c$. All electron flows 
		around potential valleys are closed thus make the system insulating.}\label{fig05}
\end{figure}

In addition, under PBC the quantum phases on $\theta=0$ line are 
symmetric about $p=p_c$. 
There are two strategies to understand this symmetry.
The first one comes from global considerations.
To begin with, a given arbitrary random scalar potential profile (with 
statistical average being zero) is denoted as $\Sigma$.
Then, following our sketch rules, we define 
$\mathbb{A}_{\Sigma\ominus}^{\uparrow(\downarrow)}(p)$ as the network composed of 
all solid closed loops around potential valleys with up(down) spin.
They are inter-connected by dashed SPTs for $p>p_c (E_F<0)$, as shown in Fig. 1c.
Since $\theta=0$, $\mathbb{A}_{\Sigma\ominus}^{\uparrow}(p)$ is decoupled from
$\mathbb{A}_{\Sigma\ominus}^{\downarrow}(p)$, although they coincide with
each other in real space.
Similarly the network including all SPT-interconnected closed loops around 
potential peaks with up(down) spin for $p'<p_c (E_F>0)$ are defined as 
$\mathbb{B}_{\Sigma\oplus}^{\uparrow(\downarrow)}(p')$ (not shown in Fig. 1). 
Also $\mathbb{B}_{\Sigma\oplus}^{\uparrow}(p')$ is unrelated to 
$\mathbb{B}_{\Sigma\oplus}^{\downarrow}(p')$ in the absence of SMT.
For any $p_1(>p_c)$, by definition we have the following mappings under PBC,
\begin{equation}\label{AB_set_correspondence_1}
\begin{array}{rcl}
p_1 &\to & 1-p_1  \\
\mathbb{A}_{\Sigma\ominus}^{\uparrow}(p_1)&\to& \mathbb{B}_{\Sigma\oplus}^{\uparrow}(1-p_1)   \\
\mathbb{A}_{\Sigma\ominus}^{\downarrow}(p_1)&\to& \mathbb{B}_{\Sigma\oplus}^{\downarrow}(1-p_1). \\ 
\end{array}
\end{equation}
On the other hand, we define $-\Sigma\equiv\overline{\Sigma}$.
Obviously, peaks (valleys) of $\overline{\Sigma}$ are valleys (peaks) of $\Sigma$,
hence S-type (S'-type) PSPs of $\overline{\Sigma}$ are S'-type (S-type) PSPs of $\Sigma$.
By symmetry, under PBC one has
\begin{equation}\label{AB_set_correspondence_2}
\mathbb{B}_{\Sigma\oplus}^{\uparrow}(1-p_1)\equiv\mathbb{A}_{\overline{\Sigma}\ominus}^{\downarrow}(p_1),   \;
\mathbb{B}_{\Sigma\oplus}^{\downarrow}(1-p_1)\equiv\mathbb{A}_{\overline{\Sigma}\ominus}^{\uparrow}(p_1). 
\end{equation}
Then the mappings in Eq. (\ref{AB_set_correspondence_1}) becomes
\begin{equation}\label{AB_set_correspondence_3}
\begin{array}{rcl}
p_1 &\to & 1-p_1  \\
\mathbb{A}_{\Sigma\ominus}^{\uparrow}(p_1)&\to& \mathbb{A}_{\overline{\Sigma}\ominus}^{\downarrow}(p_1)   \\
\mathbb{A}_{\Sigma\ominus}^{\downarrow}(p_1)&\to& \mathbb{A}_{\overline{\Sigma}\ominus}^{\uparrow}(p_1). \\ 
\end{array}
\end{equation}
Note that both $\Sigma$ and $\overline{\Sigma}$ are examples of 
``random scalar potential with zero statistical average".
Then naturally $G_{\mathrm{2T}}^{\mathrm{PBC}}$ and $\Lambda^{\mathrm{PBC}}$ 
are both statistically symmetric about $p=p_c$.

The second strategy focuses locally on each PSP, in which  
$T_0$ and $T'_0$ (kernels of transfer matrices $T_S$ and $T_{S'}$)
are the main concern.
At $\theta=0$, for an arbitrary $p (0<p<1)$,
Eqs. (\ref{Transfer_matrix_T_0}) and (\ref{Transfer_matrix_Tp_0})
provide
\begin{equation}\label{Theta_0_T0_T0p}
\begin{array}{rcl}
T_0(p,0) &= & \frac{1}{\sqrt{1-p}}\sigma_0\otimes\sigma_0-\sqrt{\frac{p}{1-p}}\sigma_x\otimes\sigma_0,  \\
T'_0(p,0) &= & \frac{1}{\sqrt{p}}\sigma_z\otimes\sigma_0-\sqrt{\frac{1-p}{p}}\mathrm{i}\sigma_y\otimes\sigma_0.   \\
\end{array}
\end{equation}
Then the following connections
\begin{equation}\label{Theta_0_T0_T0p_connection}
\begin{array}{rcl}
T_0(1-p,0)  &= & [\sigma_z\otimes\sigma_0]\cdot T'_0(p,0)  \\
T'_0(1-p,0) &= & [\sigma_z\otimes\sigma_0]\cdot T_0(p,0)   \\
\end{array}
\end{equation}
hold.
A possible misunderstanding must be clarified here. 
The ``$p\leftrightarrow 1-p$" mapping does not change the random scalar 
potential profile. S-type (S'-type) PSPs are always S-type (S'-type).
What it really changes is the Fermi level of this system, i.e.
from ``$E_F(p)$" to ``$-E_F(p)$" due to Eq. (\ref{p_expression_1}), 
since we have fixed the energy reference point to be zero.
Under our sketch rules, at a S-type PSP, for ``$p(>p_c)$", the electron
flows are shown in Fig. 1c. For ``$1-p$", the valley-peak-distribution 
is unchanged but the electron flows change to those depicted in Fig. 1d.
Now the PSP is still S-type and only its scattering matrix 
takes a similar mathematical format as a S'-type PSP.
Bearing this in mind, the connection (\ref{Theta_0_T0_T0p_connection}) 
actually means at a certain PSP, the transfer matrix at ``$p$" in an original
closed equipotential loop surrounding a potential valley (peak) is 
mathematically related to the transfer matrix at ``$1-p$" in a mapped 
loop around a potential peak (valley).
This is exactly what Eq. (\ref{AB_set_correspondence_1}) tells us.
Therefore mathematically S-type and S'-type PSPs exchange their roles
in constructing the total transfer matrix. 
Hence under PBC, the total transfer matrix
is unchanged, resulting in the symmetry about $p=p_c$.

\subsubsection{IV.A.2 Asymmetry about $p=p_c$ when $\theta>0$}
When $\theta>0$ (SMT appears), an intermediate NM phase emerges
between the two NI phases (PBC) or ``QSH+NI" phases (RBC),
as a manifestation of Wigner-Dyson symplectic ensembles.
Numerical data in Fig. 4a clearly show that the phase diagram is 
asymmetric about $p=p_c$ line.
This can also be explained by the global and local strategies
introduced in the above subsection.

From the global strategy, the definitions of 
$\mathbb{A}_{\Sigma\ominus}^{\uparrow(\downarrow)}(p)$ and 
$\mathbb{B}_{\Sigma\oplus}^{\uparrow(\downarrow)}(p')$ are unchanged.
However, now $\mathbb{A}_{\Sigma\ominus}^{\uparrow}(p)$ is coupled 
with $\mathbb{A}_{\Sigma\ominus}^{\downarrow}(p)$ via SMT.
The situation is similar for 
$\mathbb{B}_{\Sigma\oplus}^{\uparrow(\downarrow)}(p')$.
The mappings in Eq. (\ref{AB_set_correspondence_1}) still hold.
But the symmetry in Eq. (\ref{AB_set_correspondence_2}) fails
due to the SMT terms. 
Hence the final mappings in Eq. (\ref{AB_set_correspondence_3}) do not 
exist, leading to the asymmetry about $p=p_c$ line when $\theta>0$.

From the local strategy, the general form of $T_0(p,\theta)$ and
$T'_0(p,\theta)$ are given in Eqs. (\ref{Transfer_matrix_T_0}) 
and (\ref{Transfer_matrix_Tp_0}).
For $0<\theta\le \pi/2$, generally $T_0(1-p,\theta)$ and
$T'_0(1-p,\theta)$ have no explicit connections with $T'_0(p,\theta)$ and
$T_0(p,\theta)$ as in Eq. (\ref{Theta_0_T0_T0p_connection}).
This also explains the asymmetry about $p=p_c$.

\subsubsection{IV.A.3 QSH phase on $p=0$}
On the vertical $p=0$ line in Fig. 4a, Eq. (\ref{Transfer_matrix_T_0})
becomes
\begin{equation}\label{p_0_T0}
T_0(0,\theta)=\sigma_0\otimes\sigma_0,
\end{equation}
which is irrelevant to $\theta$, meaning that the SMT has no 
effects on ``left-to-right" transfer of electron flows.
However from Eq. (\ref{Transfer_matrix_Tp_0}), $T'_0(0,\theta)$ provides 
singularity since $r=\sqrt{p}=0$.
This is due to the fact that when $p=0$, at S'-type PSPs in the bulk, 
terminals on the left-hand side are decoupled from
those on the right-hand side, thus have no contributions to 
left-to-right transfer.
All electron current loops around potential peaks then become completely closed.
Under RBC, at boundary S'-type PSPs, the completely reflecting of electron
flows results in dissipationless edge modes thus make the system fall into QSH phase.

\subsubsection{IV.A.4 One-to-one mapping between $p=1$ and $\theta=\frac{\pi}{2}$ lines}
On $p=1$ line, one has
\begin{equation}\label{p_1_T0_T0p}
\begin{array}{rcl}
T_0(1,\theta) &= & \frac{\cos\theta}{\sin\theta}\sigma_0\otimes\mathrm{i}\sigma_y-\frac{1}{\sin\theta}\sigma_x\otimes\mathrm{i}\sigma_y,  \\
T'_0(1,\theta) &= & \cos\theta\sigma_z\otimes\sigma_0+\sin\theta\mathrm{i}\sigma_y\otimes\mathrm{i}\sigma_y.   \\
\end{array}
\end{equation}
While on $\theta=\pi/2$ line, the counterparts are
\begin{equation}\label{theta_half_pi_T0_T0p}
\begin{array}{rcl}
T_0(p,\frac{\pi}{2}) &= & \sqrt{1-p}\sigma_0\otimes\sigma_0-\sqrt{p}\sigma_x\otimes\mathrm{i}\sigma_y,  \\
T'_0(p,\frac{\pi}{2}) &= & -\sqrt{\frac{1-p}{p}}\sigma_z\otimes\mathrm{i}\sigma_y+\frac{1}{\sqrt{p}}\mathrm{i}\sigma_y\otimes\mathrm{i}\sigma_y.   \\
\end{array}
\end{equation}
If we perform the bijection
\begin{equation}\label{p_1_theta_half_pi_bijection}
\sqrt{p} \leftrightarrow  \sin\theta, \quad \sqrt{1-p} \leftrightarrow  \cos\theta
\end{equation}
between the two line segments 
$\left\{p=1,\theta\in (0,\frac{\pi}{2}]\right\}$ and
$\left\{\theta=\frac{\pi}{2},p\in (0,1]\right\}$,
then the following connections
\begin{equation}\label{p_1_theta_half_pi_T0_T0p_connection}
\begin{array}{rcl}
T_0(p,\frac{\pi}{2}) &= & T'_0(1,\theta)\cdot[-\sigma_z\otimes\sigma_0]  \\
T'_0(p,\frac{\pi}{2}) &= & [-\sigma_z\otimes\sigma_0]\cdot T_0(1,\theta)   \\
\end{array}
\end{equation}
hold. Note although the unitary matrix ``$-\sigma_z\otimes\sigma_0$" lies
on different sides, its $\pi$-phases (originated from diagonal ``$-1$" elements)
can be absorbed into phase matrices $\mathcal{U}_1$ and $\mathcal{U}'_1$,
thus do not affect the mathematical role-reversal of S- and S'-type PSPs 
under bijection (\ref{p_1_theta_half_pi_bijection}). 
Then $p=1$ and $\theta=\pi/2$ lines are equivalent and both fall
into NI phase under PBC.
Under RBC, on $\theta=\pi/2$ line dissipationless edge modes appear 
at boundary S'-type PSPs thus make the system fall into QSH phase.
While for $p=1$ line, similar to Fig. 5b, closed electron-flow loops
around potential valleys turn the system to NI state.

At last, at the phase point $(p,\theta)=(1,\frac{\pi}{2})$,
which is the cross point of the above two line segments, one has
\begin{equation}\label{p_1_theta_half_pi_T0_eq_T0p}
T_0(1,\frac{\pi}{2})=T'_0(1,\frac{\pi}{2}) = -\sigma_x\otimes\mathrm{i}\sigma_y.
\end{equation}
The completely diagonal transfer matrices fully mix the up and down
spins and meantime greatly enhance the itinerant range of electrons. 
Thus at this very point, the system becomes metallic.

\subsection{IV.B Mapping to phase diagram of $Z_2$-QNM}
In fact, we can map our phase diagram (Fig. 4a) to a more symmetric one.
However, before do that, it is interesting to point out that our 
phase diagram has close connection with that from the existing 
$Z_2$-QNM [see Fig. 8 and Fig. 11 in Ref. \cite{Obuse_2007_prb}]:
under PBC, they are symmetric about the vertical $p=p_c$ line.

\begin{figure*}[htbp]
	\centering
	\scalebox{0.75}[0.75]{\includegraphics[angle=0]{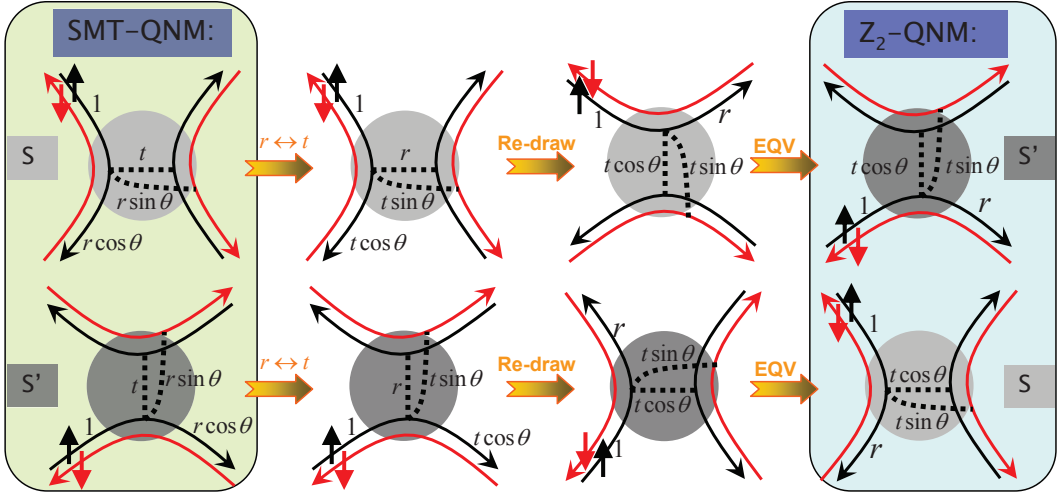}}
	\caption{(Color online) Illustration of the mapping from SMT-QNM to
		$Z_2$-QNM in the case of $p>p_c (r>t)$. There are mainly three 
		steps: (s1) $p\to 1-p$, or equivalently $t\leftrightarrow r$;
		(s2) redraw the electron flows based on our sketch rules;
		(s3) exchange up and down spins.
		Then the S-type (S'-type) PSPs in SMT-QNM has the same electron-flow 
		structure as the S'-type (S-type) PSPs in $Z_2$-QNM.}\label{fig06}
\end{figure*}

The reason is straightforward. 
By mapping the horizontal axis ``$x$" in Ref. \cite{Obuse_2007_prb}
to the counterpart in this work ``$p$" through $p=\tanh^2 x$, we
rewrite their Eq. (2.3) in terms of ``$r$" and ``$t$" as
\begin{equation}\label{Transfer_matrix_T_0_Z2}
\begin{array}{rcl}
T_0^{Z_2}&=&\frac{1}{t}\left[\sigma_0\otimes (T_0^{Z_2})^{\mathrm{d}}+\sigma_x\otimes (T_0^{Z_2})^{\mathrm{od}}\right],    \\
(T_0^{Z_2})^{\mathrm{d}}&=&\left(
\begin{matrix}
\cos\theta & r\sin\theta  \\
-r\sin\theta & \cos\theta  \\
\end{matrix}
\right),   \\
(T_0^{Z_2})^{\mathrm{od}}&=&\left(
\begin{matrix}
-r\cos\theta & -\sin\theta  \\
\sin\theta & -r\cos\theta  \\
\end{matrix}
\right),   \\
\end{array}
\end{equation}
and 
\begin{equation}\label{Transfer_matrix_Tp_0_Z2}
\begin{array}{rcl}
T'_0{}^{Z_2}&=&\frac{1}{1-t^2\cos^2\theta}\left[\sigma_z\otimes (T'_0{}^{Z_2})^{\mathrm{d}}+\mathrm{i}\sigma_y\otimes (T'_0{}^{Z_2})^{\mathrm{od}}\right],    \\
(T'_0{}^{Z_2})^{\mathrm{d}}&=&\left(
\begin{matrix}
r & -t^2\sin\theta\cos\theta  \\
t^2\sin\theta\cos\theta & r  \\
\end{matrix}
\right),   \\
(T'_0{}^{Z_2})^{\mathrm{od}}&=&\left(
\begin{matrix}
-rt\cos\theta & t\sin\theta  \\
-t\sin\theta & -rt\cos\theta  \\
\end{matrix}
\right).   \\
\end{array}
\end{equation}
By comparing them with the transfer matrix kernels $T_0$ and $T'_0$ 
of our SMT-QNM, we have the following connections
\begin{equation}\label{Z2_SMT_connection}
\begin{array}{rcl}
T_0^{Z_2}(p,\theta)  &= & [\sigma_z\otimes\sigma_z]\cdot T'_0(1-p,\theta)\cdot [\sigma_0\otimes\sigma_z], \\
T'_0{}^{Z_2}(p,\theta) &= & [\sigma_z\otimes\sigma_z]\cdot T_0(1-p,\theta)\cdot [\sigma_0\otimes\sigma_z],   \\
\end{array}
\end{equation}
which are quite similar to Eq. (\ref{Theta_0_T0_T0p_connection}).
Therefore, similar discussions as in the end of Sec. IV.A.1 can be performed. 

In Fig. 6 we illustrate a typical mapping starting from the SMT-QNM with 
$p>p_c (r>t)$. 
The main procedure is: (s1) $p\to 1-p$, or equivalently exchange $t$ and $r$;
(s2) following our sketch rules, the electron flows are redrawn;
(s3) by exchanging up and down spins,
the S-type (S'-type) PSPs in SMT-QNM has the same electron-flow structure
as the S'-type (S-type) PSPs in $Z_2$-QNM. Then it is understandable 
that under PBC by performing a mirror-symmetry operation on our phase diagram (Fig. 4a)
about $p=p_c$ line, one gets the phase diagram of the $Z_2$-QNM.
Note that in this mapping only the electron-flows are converted.
The potential valleys and peaks are unchanged.

Based on this result, the critical exponent and normalized localization
length at phase transitions should be the same as those from $Z_2$-QNM.
This is confirmed by numerical calculations within error permissibility.
To save space, we do not show this part of our data here. 
However, this close connection should not downgrade the significance of
SMT-QNM constructed in this work.
First, in our SMT-QNM, SMT process is an additional 
tunneling channel and does not take probability away from
the existing SPT channel, which is a more physical assumption.
Second, the symmetry about $p=p_c$ line between these two phase diagrams 
indicates a possible way to check which network model
provides better description to real 2D-DSEGs.
From Eq. (\ref{p_expression_1}), $p$ is directly related to system
Fermi level. By sweeping the Fermi level and check out the quantum phase
a 2D-DSEG falls in, experimentally one can make reasonable judgment.
Third, as will be shown next, the phase diagram of
SMT-QNM can be topologically transformed to a symmetric one which is 
highly similar to the phase diagram of disordered 3D weak TIs.
This enriches the possible applications of our 2D SMT-QNM.

\subsection{IV.C Mapping to a symmetric phase diagram}
The asymmetry of phase diagram in the original $(p,\theta)$ 
phase space is unfavorable for a deep understanding of ATs in 2D-DSEGs.
Fortunately, its features summarized in Sec.VI.A provide us enough 
information to topologically transform it to a completely symmetric one.
Mathematically, the following mapping
\begin{equation}\label{p_theta_to_X_Y}
X=\frac{p\cos^2\theta}{1-p\sin^2\theta},\quad Y=p\sin^2\theta
\end{equation}
perfectly achieves this target:

\noindent
(a) The original phase space ``$\Omega_1$" [see Eq. (\ref{Phase_space_p_theta})]
is mapped to a new phase space
\begin{equation}\label{Phase_space_X_Y}
\Omega_2\equiv \left\{(X,Y)|0\le X\le1,0\le Y \le1\right\};
\end{equation}

\noindent
(b) The original $\theta=0$ line is mapped to $Y=0$ line with the one-to-one
correspondence $p\leftrightarrow X$, hence $Y=0$ line is symmetric about 
the vertical line $X=X_c(\equiv p_c)$;

\noindent
(c) The original $p=0$ line shrinks to a single point $(X,Y)=(0,0)$;

\noindent
(d) The metallic phase-point $(p,\theta)=(1,\frac{\pi}{2})$ stretches
itself to $Y=1$ line, which is indeed a singularity of the mapping in 
Eq. (\ref{p_theta_to_X_Y}); 

\noindent
(e) The $\theta=\frac{\pi}{2}$ and $p=1$ lines are mapped to $X=0$ and $X=1$ 
lines, respectively. The combination of mappings in 
Eqs. (\ref{p_1_theta_half_pi_bijection}) and (\ref{p_theta_to_X_Y}) generates
a one-to-one correspondence of phase points on $X=0$ and $X=1$ lines with the same $Y$.
Therefore these two lines are completely symmetric about $X=X_c$ line.

\noindent
(f) For an arbitrary point $(X,Y)$ in the interior region of $\Omega_2$,
one has
\begin{equation}\label{T_0_X_Y}
\begin{array}{rcl}
T_0(X,Y)&=&\frac{1}{1-X+XY}\left[\sigma_0\otimes \Delta^{\mathrm{d}}+\sigma_x\otimes \Delta^{\mathrm{od}}\right],    \\
\Delta^{\mathrm{d}}&=&\sqrt{(1-X)(1-Y)}\sigma_0+\sqrt{XY(1-Y)}\mathrm{i}\sigma_y,   \\
\Delta^{\mathrm{od}}&=&-(1-Y)\sqrt{X(1-X)}\sigma_0-\sqrt{Y}\mathrm{i}\sigma_y,   \\
\end{array}
\end{equation}
and
\begin{equation}\label{Tp_0_X_Y}
\begin{array}{rcl}
T'_0(X,Y)&=&\frac{1}{X+Y-XY}\left[\sigma_z\otimes \Delta'{}^{\mathrm{d}}+\mathrm{i}\sigma_y\otimes \Delta'{}^{\mathrm{od}}\right],    \\
\Delta'{}^{\mathrm{d}}&=&\sqrt{X(1-Y)}\sigma_0-\sqrt{(1-X)Y(1-Y)}\mathrm{i}\sigma_y,   \\
\Delta'{}^{\mathrm{od}}&=&-(1-Y)\sqrt{X(1-X)}\sigma_0+\sqrt{Y}\mathrm{i}\sigma_y.   \\
\end{array}
\end{equation}
Then the following connections
\begin{equation}\label{Phase_diagram_symmetry_in_XY}
\begin{array}{rcl}
T_0(1-X,Y)  &= & [\sigma_z\otimes\sigma_z]\cdot T'_0(X,Y)\cdot [\sigma_0\otimes\sigma_z]  \\
T'_0(1-X,Y) &= & [\sigma_z\otimes\sigma_z]\cdot T_0(X,Y)\cdot [\sigma_0\otimes\sigma_z]   \\
\end{array}
\end{equation}
indicate the symmetry of mapped phase diagram about $X=X_c$ line in $\Omega_2$.

Following the mapping in Eq. (\ref{p_theta_to_X_Y}), we transform
the phase diagram in $\Omega_1$ into the one in $\Omega_2$ which is plotted
in Fig. 4b.
Obviously, the new phase diagram looks better. However, it is not 
completely symmetric about $X=X_c$ due to the finite-size effect
during our calculation, since we only perform
calculations on normalized localization length to $W=2^5$ 
limited by our existing computing capability.
It is expected that when $W$ is sufficient large, the symmetry
in $\Omega_2$ should be more apparent.

The results in this subsection have several potential applications. 
First, phase boundaries in $\Omega_1$ can be double-checked through 
the mapping (\ref{p_theta_to_X_Y}) and its inversion,
since in $\Omega_2$ phase boundaries should be symmetric about $X=X_c$.
Second, the narrow and long NI (or QSH) phase in the close vicinity of 
$(p,\theta)=(1,\frac{\pi}{2})$, which is hard to precisely determined
due to strong symmetry-crossover effects, is enlarged a bit in $\Omega_2$. 
This should be helpful for better determination of NM-NI (QSH) boundaries.

\subsection{IV.D Connection with disordered 3D weak TIs}
In addition, the new phase diagram (Fig. 4b) shows apparent similarity
with that of disordered 3D weak TIs (see Fig. 1 in Ref.\cite{Moore_PRL_2012}),
indicating a close connection between 2D-DSEGs described by our SMT-QNM 
and the helical surface modes of 3D weak TIs under scalar disorder 
potentials respecting TRS.
Comparing our 2D Dirac Hamiltonian 
[see Eq. (\ref{Dirac_H_final_valley_description}) in this work] and the effective
Hamiltonian in Ref.\cite{Moore_PRL_2012} [see Eqs. (1)-(3) therein], 
the energy gap of the system is $2|m|$ with $m=1-\frac{p}{p_c}=1-\frac{X}{X_c}$.
For clean limit, $V_{00}=A_0=0$ meaning on $Y=0$ line in Fig. 4b, the 
intermediate metallic region shrinks to a single critical point $X=X_c$.
In the presence of disorder which couples the two Dirac fermions (with mass $\pm m$)
with strength $V_{x0}=\sqrt{2}\theta$, direct transitions between
the insulating phases (NI and QSH) are forbidden due to the stability of
the symplectic metal, 
which results in the finite width of intermediate metallic phase.
In addition, the disorder strength 
\begin{equation}\label{g_sin_theta_square}
g\sim |V_{x0}|^2\sim \theta^2\sim\sin^2\theta\propto Y
\end{equation}
in the vicinity of the critical point $(X,Y)=(X_c,0)$.
All these correspondences confirm the close connection we mentioned
at the beginning of this section.
This implies the possible application of our SMT-QNM on
investigations of disordered helical surface modes of 3D weak TIs.
For $Z_2$-QNM, similar works have been done systematically\cite{Obuse_2014_prb}.
For our SMT-QNM, this is an interesting direction but out of the scope
of this work.

\section{V. Quantum phases and phase transitions in TRS-breaking SMT-QNM}
The TRS-preserving SMT-QNM introduced above can be downgraded to
the counterpart which still preserves TRS at PSPs but breaks it in the links
between PSPs.
Physically, this corresponds to 2D-DSEGs with TRS-breaking 
(usually called magnetic) isotropic impurities.
These impurities inevitably affect the random potential profile, however
will not create PSPs at their very locations due to the isotropic nature, 
thus can be described by the TRS-breaking SMT-QNM.
In these systems, spin-flip backscattering on each link between PSPs 
emerges thus destroys the original Krammer's doublet.
For modelization, this can be simply realized by neglecting the 
``phase pairing rule" in Eq. (\ref{Random_phase_matrice_2}),
meanwhile leaving the rest unchanged.
Here we briefly summarize our data and provide reasonable explanations.

\subsection{V.A Phase diagram}
Now the system falls into Wigner-Dyson unitary class 
(TRS fails, regardless of SRS) and generally no intermediate NM
phase exists. This is confirmed by finite-size analysis on
$G_{\mathrm{2T}}$ and $\Lambda$.
The resulting phase diagram is plotted in Fig. 7 and totally different 
from the TRS-preserving SMT-QNM.
Our data show that QSH state only survives on the line 
segment $\{0\le p<0.5,\theta=0\}$ if we choose S'-type PSPs as boundary nodes.
When SMT emerges, there is no intermediate NM phase. 
In the entire phase space $\Omega_1$, NI phase dominates.
\begin{figure}[htbp]
	\centering
	\scalebox{1.0}[1.0]{\includegraphics[angle=0]{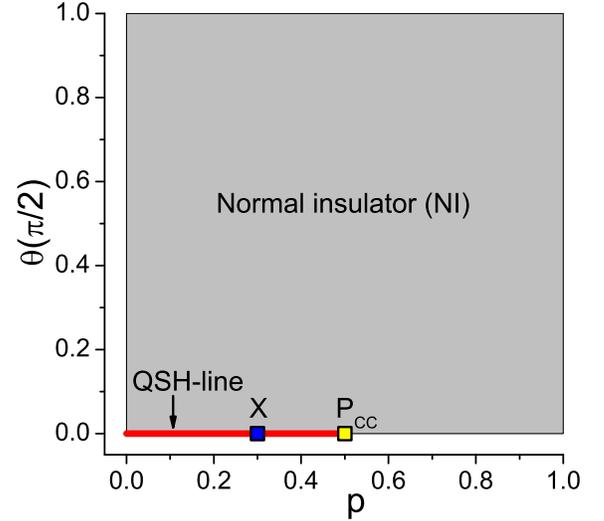}}
	\caption{(Color online) Phase diagram of the TRS-breaking SMT-QNM under RBC
		with boundary nodes being S'-type PSPs. NI phase (gray area)
		dominates in the phase space $\Omega_1$ and QSH phase (red solid line) 
		only survives on the line segment $\{0\le p<0.5,\theta=0\}$. 
		The point $P_{\mathrm{CC}}$ is the CC-RNM critical point $(p_c,0)$.
		The point $X$ is a typical QSH phase point with the coordinate:
		$X:(0.3,0)$.}\label{fig07}
\end{figure}

\subsection{V.B The NI phase}
The TRS-breaking in links connecting PSPs will turns both NM and QSH 
phases (except for the segment on $\theta=0$ line) into NI phase, 
which is the typical behavior of unitary ensembles.
To check for this, first we perform numerical calculations of $G_\mathrm{2T}$
under RBC for enough dense grid of the phase space $\Omega_1$.
For all phase points, the network size $W(=L)$ increases from $2^2$ to $2^9$.
Further enlargement of $W$ is out of our computing capability.
The sample number is always 128, which is enough to provide sufficient 
small error.
To save space, we summarize the main features and present
typical data, if necessary. 
First, for all phase points in $\Omega_1$, 
$\left\langle G_{\mathrm{2T}}^{\mathrm{RBC}} \right\rangle$ 
is smaller than 1 and decrease with $W$ 
for sufficient large $W$ without sign of convergence.
Obviously this can not be QSH state.
In addition, we have known that in NM phase (if exists) of 
systems with unitary symmetry,
$\left\langle G_{\mathrm{2T}}^{\mathrm{RBC}} \right\rangle$
should converge to the Boltzmann conductance
$G_0(\gg 1)$\cite{Hikami_1980}.
Hence our data clearly show that the system falls into neither QSH nor
unitary metallic phase. The only possibility is the NI phase.

\begin{figure}[htbp]
	\centering
	\scalebox{1.1}[1.1]{\includegraphics[angle=0]{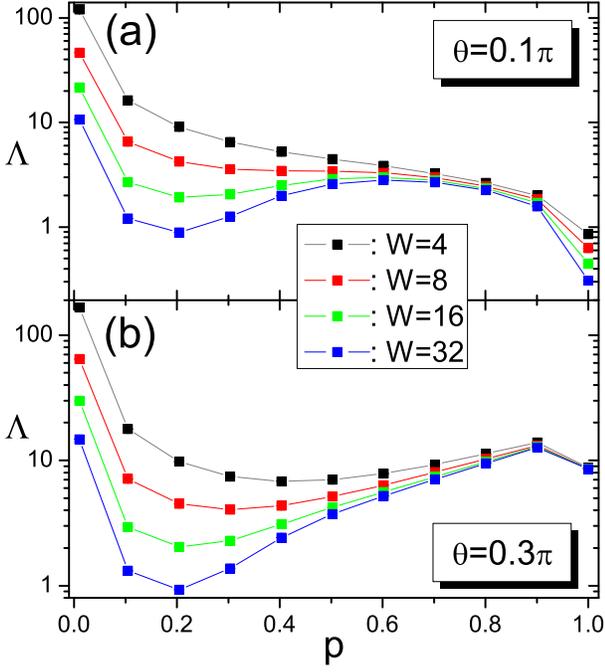}}
	\caption{(Color online) Finite-size calculations of the reduced 
		localization length $\Lambda$ along (a) $\theta=0.1\pi$ 
		and (b) $\theta=0.3\pi$ with RBC in transverse direction.
		The relative standard error of the first LE is
		set to be $1\%$ and the sample number is 32.
	    All error bars are smaller than data symbols.}\label{fig08}
\end{figure}

Next, the finite-size calculations of the normalized localization
length $\Lambda$ for $\theta=0.1\pi$ and $\theta=0.3\pi$ are performed 
and the data are plotted in Fig. 8.
In these calculates, RBC in transverse direction is imposed and the 
sample number is chosen to be 32.
In addition, the relative standard error of the first LE is set to be $1\%$.
The numerics clearly shows that for both SMT strengths, 
$\Lambda$ always decreases with the network width $W$.
This confirms that along $\theta=0.1\pi$ and $\theta=0.3\pi$ lines
in $\Omega_1$, the system falls into NI phase.
Similar calculations have been performed for other nonzero $\theta$ values.
All results support our conclusion that NI phase fills up $\Omega_1$.

\subsection{V.C Direct transition from QSH to NI phases}
The QSH phase on line segment $\{0\le p<0.5,\theta=0\}$
is absolutely unstable to SMT.
This means no matter how small the SMT is, the QSH state will 
be destroyed completely.
To see it, the point X ($p=0.3,\theta=0$) 
in Fig. 7 is selected as an example.
We set $W=L$ and vary $W$ from $2^1$ to $2^9$.
In the close vicinity of point X,
$\left\langle G_{\mathrm{2T}}^{\mathrm{RBC}} \right\rangle$
and the corresponding error of 128 independent configurations are 
calculated and plotted in Fig. 9.
For point X, numerical data (hollow squares in Fig. 9) show that 
when the system size increases to $W=2^9$,
$\left\langle G_{\mathrm{2T}}^{\mathrm{RBC}} \right\rangle$ 
approaches the quantized value 2, 
with the standard error as small as $5.6\times 10^{-9}$. 
This validates that point X belongs to the QSH phase.
Next we perform calculations for $\theta=0.01$. 
The result is shown in Fig. 9 by solid magenta squares. 
As the system size gets larger, 
$\left\langle G_{\mathrm{2T}}^{\mathrm{RBC}} \right\rangle$ 
falls to $10^{-3}$ or even smaller.
We then gradually approach the point X by decreasing $\theta$ 
by an order of magnitude and calculate the corresponding
$\left\langle G_{\mathrm{2T}}^{\mathrm{RBC}} \right\rangle$ 
until $\theta$ reaches $1.0\times 10^{-6}$.
The results are plotted in Fig. 9, showing that 
as $\theta$ decreases, the deviation of 
$\left\langle G_{\mathrm{2T}}^{\mathrm{RBC}} \right\rangle$
from quantized value 2 
gets weaker at $W=L=2^9$. 
However, it always exists and has no sign of convergence.
Even for $q=1.0\times 10^{-6}$ (solid black squares), 
if the system size is further increased to $2^{10}$, 
$\left\langle G_{\mathrm{2T}}^{\mathrm{RBC}} \right\rangle$ deviates from 2 evidently.
Limited by computing capability, we can not perform calculations
to the system size at which 
$\left\langle G_{\mathrm{2T}}^{\mathrm{RBC}} \right\rangle$ 
falls to zero.
However, the data in Fig. 9 clearly imply that 
\emph{QSH state can not survive when SMT emerges, no matter how small it is.}

\begin{figure}[htbp]
	\centering
	\scalebox{0.85}[0.85]{\includegraphics[angle=0]{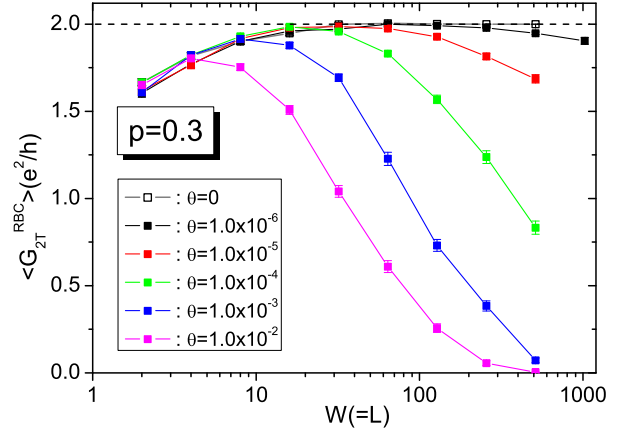}}
	\caption{(Color online) Evolution of 
		$\left\langle G_{\mathrm{2T}}^{\mathrm{RBC}} \right\rangle$ 
		around point X in Fig. 7 when SMT ($\theta$) appears. 
		The system size $W(=L)$ increases from $2^1$ to $2^9$ for
		$q=0,10^{-2},10^{-3},10^{-4},10^{-5}$ and $10^{-6}$ with 
		the sample numbers all equal to 128. 
		In particular, for $q=10^{-6}$, the system size
		increases further to $2^{10}$ with the sample number being 16.}\label{fig09}
\end{figure}

For cross validation, we also perform finite-size calculations for
the normalized localization length $\Lambda$, in which RBC is adopted
and the sample number is 16. In addition, the relative standard
error of the first LE is $1\%$, leading to the network length 
$L\sim10^6$.  The network width
$W$ increases from $2^2$ to $2^5$ and the SMT strength $\theta$ 
varies from $0$ to $0.01$ with the step $\mathrm{d}\theta=0.001$.
The resulting data are plotted in Fig. 10.
It is clear that $\Lambda$ for $\theta=0$ always increases with $W$.
This comes from the dissipationless edge modes and confirms the fact that 
point X belongs to QSH phase.
On the other hand, when $\theta\ge \mathrm{d}\theta$, $\Lambda$
eventually decreases as $W$ increases to $2^5$. 
This validates the conclusion based on data from
$\left\langle G_{\mathrm{2T}}^{\mathrm{RBC}} \right\rangle$
that $(p=0.3,\theta\ge \mathrm{d}\theta)$ falls into NI phase.
Further increase of $W$ and decrease of $\mathrm{d}\theta$ are beyond
our present computing capability.
However, the data in Fig. 10 already provide enough cross-validation
evidences of the absolute instability of QSH state.

\begin{figure}[htbp]
	\centering
	\scalebox{0.85}[0.85]{\includegraphics[angle=0]{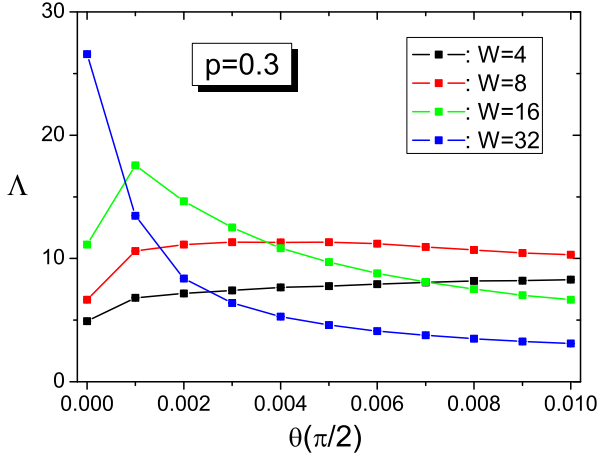}}
	\caption{(Color online) Evolution of $\Lambda$ 
		around point X in Fig. 7 when SMT ($\theta$) emerges. 
		The network width $W$ increases from $2^2$ to $2^5$. 
		The sample numbers are all 16 and the relative standard 
		error of the first LE is $1\%$. Error bars are smaller 
		than data symbols in all cases.}\label{fig10}
\end{figure}

This result can be understood by the physical process sketched in Fig. 11.
Initially the system is in QSH phase, i.e. the situation depicted in Fig. 5a.
For simplicity, we take the part close to the upper edge as an example and 
redraw it in Fig. 11.
We focus on an electron with up spin propagating in the dissipationless 
left-to-right edge channel. Suppose at some moment, the electron is at 
point A which is set as the starting point. 
When SMT is absent, the electron at most tunnels into the closed loops with
up spins via SPT and can not fall into trajectories associated with down spins.
Hence the electron will never be backscattered into the right-to-left edge 
channel with down spin at the upper edge.
On the other hand, the backscattering into the right-to-left edge channel 
with up spin at the lower edge (not depicted in Fig. 11) by means of 
multi-SPTs through closed spin-up loops will be suppressed when the network 
is wide enough since this is a high-order process.
When SMT emerges, the situation is completely different.
When a spin-up electron propagates from point A and reaches point B, 
SMT at this S-type PSP allows it to tunnel into the closed loop 
associated with down spins (point C on the red loop). 
After circling this loop (C to D to E), the electron comes back to this PSP 
and tunnels into the right-to-left edge channel (point F) with down spin 
via SPT and then go to point G and even leftward.
Now we realize a spin-flip backscattering event
(A$\rightarrow$B$\rightarrow$C$\rightarrow$D$\rightarrow$E$\rightarrow$F$\rightarrow$G)
which includes only one step of SMT.
Therefore this process is not a high-order one and should take effect 
as long as SMT appears.
Combing with the fact that a number of S-type PSPs distribute along 
the upper edge, it is understandable that the QSH state should be absolutely 
unstable with respect to SMT.
Therefore,\emph{ the QSH line segment acts as a critical line rather than a phase boundary}.
Hence the critical exponent for this direct transition can hardly
be extracted out using the standard finite-size scaling procedure\cite{Slevin_prl_1999}.
\begin{figure}[htbp]
	\centering
	\scalebox{0.5}[0.5]{\includegraphics[angle=0]{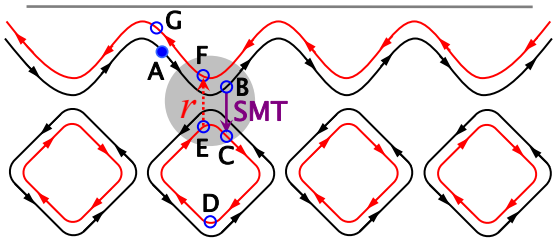}}
	\caption{(Color online) Graphic description of the spin-flip
		backscattering process induced by SMT. 
		The light gray circle indicates a S-type PSP. With the help of a
		closed loop around potential peaks, the SMT realizes a ``A$\rightarrow$B$\rightarrow$C$\rightarrow$D$\rightarrow$E$\rightarrow$F$\rightarrow$G" 
		spin-flip backscattering thus completely destroys the QSH state. }\label{fig11}
\end{figure}

\subsection{V.D Revisit of the CC-RNM critical point $(p_c,0)$}
In the end of this section, we turn back to the CC critical point
$(p,\theta)=(p_c,0)$, where the network decouples into two
copies of CC-RNM with opposite chiralities.
As mentioned above, the infinitesimal Migdal-Kadanoff 
transformation for real-space renormalization of CC-RNM
provides that it is the quantum critical point that separates
two insulating (NI and QSH) phases in the bulk.
However, \emph{``whether or not a NM phase exists 
	in a finite range around the quantum critical point $p=p_c$ 
	along $\theta=0$ line}" is still a controversial issue.
Here we present numerical data within error permissibility
and within the scope of our computing capability
to give a reasonable estimation about the width of
this NM phase, if exists.

\begin{figure}[htbp]
	\centering
	\scalebox{0.95}[0.95]{\includegraphics[angle=0]{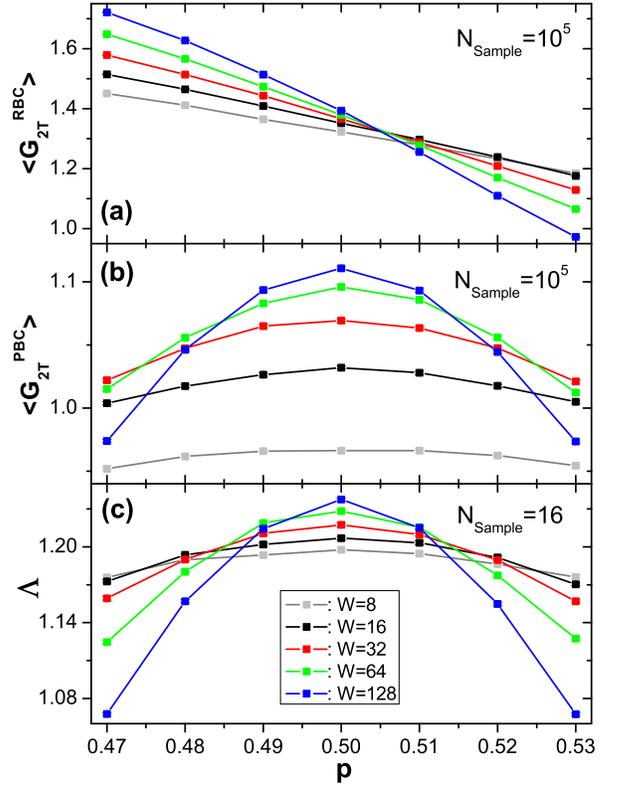}}
	\caption{(Color online) Finite-size calculations of the two-terminal
		conductance $\left\langle G_{\mathrm{2T}} \right\rangle$ under 
		RBC (a) and PBC (b), as well as the normalized 
		localization length $\Lambda$ under PBC (c) around $p=p_c$ when SMT 
		is absent ($\theta=0$). The sample numbers in (a) and (b) are $10^5$.
		For $\Lambda$ in (c), PBC is imposed 
		to eliminate the effects of dissipationless edge modes in QSH phase. 
		The relative standard error of the first LE is $0.5\%$ 
		and the sample number is 16. All error bars are smaller than data
		symbols.}\label{fig12}
\end{figure}

To begin with, we calculate $\left\langle G_{\mathrm{2T}}\right\rangle$ under both 
RBC and PBC in the vicinity of $p=p_c$ along $\theta=0$ line.
The results are shown in Fig. 12a and 12b, respectively.
In these calculations, $W(=L)$ varies from $2^3$ to $W_{\mathrm{max}}=2^7$
and the step of $p$ is $\mathrm{d}p=0.01$.
Meantime, $N=10^5$ independent samples are generated for acceptable averages.
Generally, $\mathrm{d}p$ is limited by our computing 
capability ($W_{\mathrm{max}}$ and $N$) hence can not be arbitrarily small.
In this sense, we can only detect the existence of NM phase at the level 
of $\mathrm{d}p$. For RBC, $\left\langle G_{\mathrm{2T}}^{\mathrm{RBC}}\right\rangle$ 
is an increasing function of $W$ when $p\le p_c$. When $p\ge p_c+\mathrm{d}p$, 
$\left\langle G_{\mathrm{2T}}^{\mathrm{RBC}}\right\rangle$ decreases for sufficient large $W$.
For PBC, $\left\langle G_{\mathrm{2T}}^{\mathrm{PBC}}\right\rangle$ is an increasing 
function of $W$ when $|p-p_c|\le \mathrm{d}p$, at least for $W\le W_{\mathrm{max}}$. 
While if $|p-p_c|\ge 2\mathrm{d}p$, 
$\left\langle G_{\mathrm{2T}}^{\mathrm{PBC}}\right\rangle$ eventually decreases 
at sufficient large $W$. These results imply that the NM phase at most appears 
in $(p_c-\mathrm{d}p,p_c+\mathrm{d}p)$, if exists.

To further check this conclusion, we calculate the normalized localization 
length $\Lambda$ and the data are plotted in Fig. 12c.
In these calculations, several points should be clarified.
First, PBC is adopted to eliminate the effect of dissipationless
edge states in QSH phase. Second, the relative standard error $\epsilon_1$ 
of the first LE is set to be $0.5\%$, which results in
the strip length $L\sim 10^7 \gg \xi_W$ for $W=2^7$. 
Third, the sample number is 16 and 
proved to be enough for sufficiently small error bars.
The data in Fig. 12c clear show that only at $p=p_c$, $\Lambda$ is an increasing 
function of the width $W$ (at least for $W\le W_{\mathrm{max}}$).
For other $p$ satisfying $|p-p_c|\ge \mathrm{d}p$, $\Lambda$ decreases for 
sufficient large strip width.
These results validate the fact that NM-phase is only possible to reside
in $|p-p_c|<\mathrm{d}p$, if exists, in the absence of SMT.
Further verifications need smaller $\mathrm{d}p$, greater network width
$M_{\mathrm{max}}$ and sample number $N$, 
which are out of our present computing capability.

\section{IV. Conclusions}
In this work, we have constructed the symplectic SMT-QNM by 
recognizing the SMT as an independent tunneling channel.
By leading-order expansion method, the 2D Dirac Hamiltonian is
extracted out from SMT-QNM in the close vicinity of CC-RNM critical point, 
with the SMT strength associating with the spin-flip coupling.
A sandwiched (QSH-NM-NI) phase diagram in original phase space $\Omega_1$ 
is then obtained by finite-size
analysis of two-terminal conductance and normalized localization length.
It is first mapped to the phase diagram of the existing $Z_2$-QNM,
and then closely related to the counterpart of disordered 3D weak TIs.
In the end, the TRS-breaking (in the links between PSPs) version of 
SMT-QNM is considered and turns out to fall into unitary class.
Its phase diagram is filled by NI phase except for a marginal line segment
still hosting QSH phase.
A direct transition from QSH to NI phases exists and is explained
by the SMT-induced spin-flip backscattering.

\section{Acknowledgement}
We thank Prof. X. R. Wang and Prof. G. Xiong for fruitful discussions.
This work is supported by the Science Foundation for The Excellent Youth Scholars of
Educational Commission of Hebei Province, China (Grant No. Y2012027) and the Natural
Science Foundation of Hebei Province, China (Grant No. A2014205080).
J.L also acknowledges the support from the National Natural Science Foundation of
China (Grant Nos. 11374088 and 11104060). B.Xi acknowledges the support
from the National Natural Science Foundation of China (Grant No. 11774300).

\end{document}